\documentclass[10pt, conference, compsocconf, pdftex]{IEEEtran}

\usepackage{mathptmx} 
\usepackage{mathtools}
\usepackage{hyperref}
\usepackage{multirow}
\usepackage{tikz}
\usepackage{pifont}
\usepackage{xcolor}
\usepackage[keeplastbox]{flushend}
\usepackage[normalem]{ulem}
\usepackage[sort,nocompress]{cite}
\usepackage[final]{microtype}

\newcommand*\circled[1]{\tikz[baseline=(char.base)]{
        \node[shape=circle,fill=black,draw,inner sep=0.1pt] (char) {\color{white}\fontfamily{phv}\selectfont\textbf{#1}};}}

\ifCLASSINFOpdf
  \usepackage{graphicx}
\else
\fi
\usepackage[linesnumbered,ruled]{algorithm2e}

\usepackage{array}

\usepackage{mdwmath}
\usepackage{mdwtab}

\usepackage{eqparbox}

\usepackage[caption=false,font=footnotesize]{subfig}
\usepackage{fixltx2e}

\usepackage{stfloats}

\usepackage{url}

\begin{document}
%
\title{COMPAQT: Compressed Waveform Memory Architecture for Scalable Qubit Control}


\author{\IEEEauthorblockN{Satvik Maurya}
\IEEEauthorblockA{University of Wisconsin-Madison\\
Madison, USA\\
smaurya@wisc.edu}
\and
\IEEEauthorblockN{Swamit Tannu}
\IEEEauthorblockA{University of Wisconsin-Madison\\
Madison, USA\\
swamit@cs.wisc.edu}
}


%


\maketitle

\begin{abstract}
On superconducting architectures, the state of a qubit is manipulated by using microwave pulses. Typically, the pulses are stored in the waveform memory and then streamed to the Digital-to-Analog Converter (DAC) to synthesize the gate operations. The waveform memory requires tens of Gigabytes per second of bandwidth to manipulate the qubit. Unfortunately, the required memory bandwidth grows linearly with the number of qubits. As a result, the bandwidth demand limits the number of qubits we can control concurrently. For example, on current RFSoCs-based qubit control platforms, we can control less than 40 qubits. In addition, the high memory bandwidth for cryogenic ASIC controllers designed to operate within a tight power budget translates to significant power dissipation, thus limiting scalability.

In this paper, we show that waveforms are highly compressible, and we leverage this property to enable a scalable and efficient microarchitecture {\em {\sc compaqt} - Compressed Waveform Memory Architecture for Qubit Control}.  
Waveform memory is read-only and {\sc compaqt} leverages this to compress waveforms at compile time and store the compressed waveform in the on-chip memory. To generate the pulse, {\sc compaqt} decompresses the waveform at runtime and then streams the decompressed waveform to the DACs. Using the hardware-efficient discrete cosine transform, {\sc compaqt} can achieve, on average, 5x increase in the waveform memory bandwidth, which can enable 5x increase in the total number of qubits controlled in an RFSoC setup. Moreover, {\sc compaqt} microarchitecture for cryogenic CMOS ASIC controllers can result in a 2.5x power reduction over uncompressed baseline. We also propose an adaptive compression scheme to further reduce the power consumed by the decompression engine, enabling up to 4x power reduction.    

Qubits are sensitive, and even a slight change in the control waveform can increase the gate error rate. We evaluate the impact of {\sc compaqt} on the gate and circuit fidelity using IBM quantum computers. We see less than 0.1\% degradation in fidelity when using {\sc compaqt}.
\end{abstract}


%
\IEEEpeerreviewmaketitle

\section{Introduction}
\label{sec:intro}

The true potential of quantum computers can be unlocked by building a quantum computer that can run error correction on thousands of qubits to enable fault tolerance.
Several industry and academic labs have built machines with more than fifty qubits~\cite{ibm_100q, Arute2019}, with plans to increase the number of qubits to thousands of qubits to demonstrate fault-tolerance~\cite{ibm1000}. However, building a large-scale quantum computer is challenging. Qubit devices are susceptible to noise and require precise control. Superconducting quantum computers use microwave pulses to perform operations on qubits. A typical quantum program that uses tens of qubits requires hundreds of such pulses. The control hardware sends these pulses through coaxial cables to qubits at cryogenic temperatures, while the control hardware works at room temperature. Today, all quantum computers employ brute-force approaches to scaling the control hardware where discrete, off-the-shelf components like Field Programmable Gate Arrays (FPGAs), Digital to Analog Converters (DACs), and Analog to Digital Converters (ADCs) are replicated depending on the number of qubits in the target quantum processor. For example, the control hardware of the Sycamore chip~\cite{Arute2019} required 200+ DACs, 9 ADCs, and 30+ FPGAs for control and readout of 53 qubits. 
As the number of qubits increases, this brute-force scaling of control hardware will become infeasible in terms of both cost and complexity.

\begin{figure}[t]
    \centering
    \includegraphics[width=\linewidth]{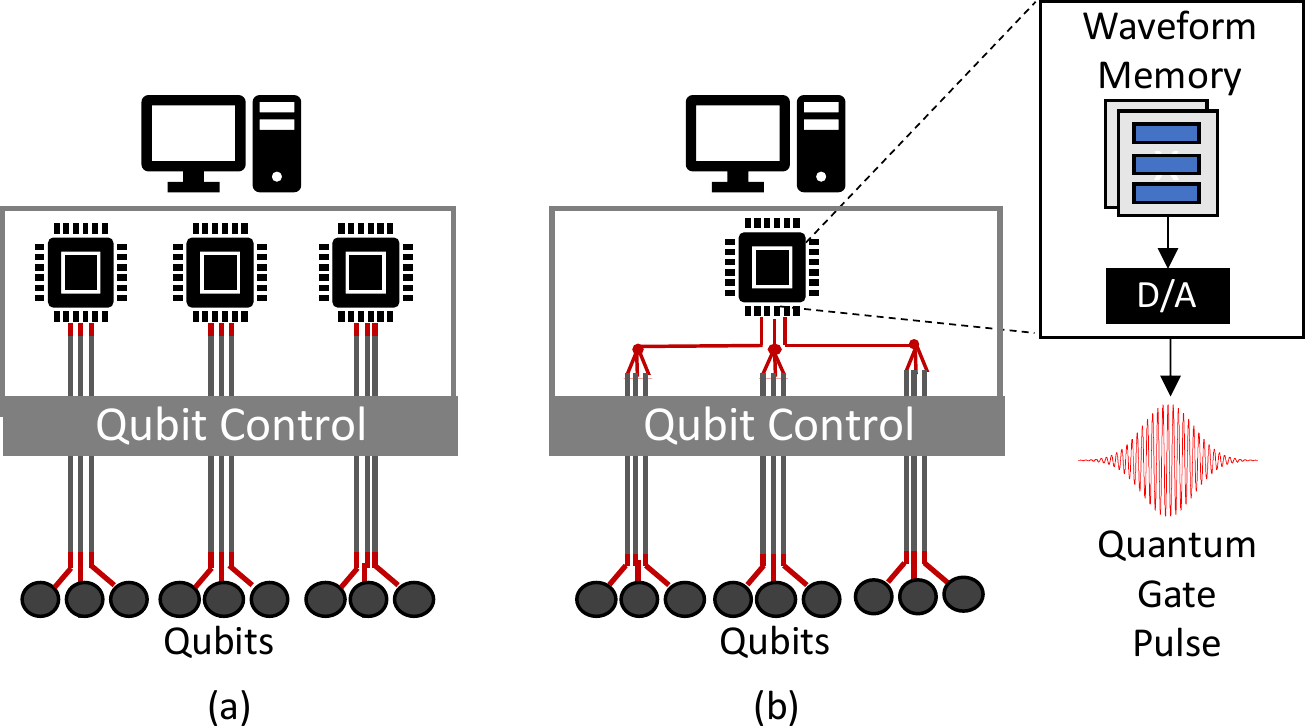}
    \caption{(a) Distributed Qubit Control with FPGAs (b) Integrated Qubit Control with RFSoCs.}
    \label{fig:fig1}
\end{figure}

\begin{figure*}[t]
    \centering
    \includegraphics[width=0.9\linewidth]{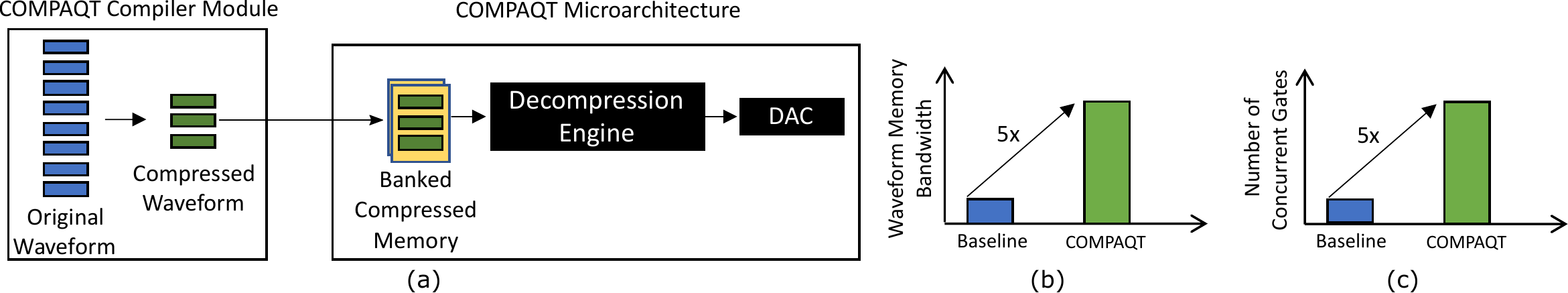}
    \caption{(a) Organization of COMPAQT (b) Waveform memory bandwidth boost with COMPAQT (c) Improvement in the number of concurrent gates that a controller can support.} 
    \label{fig:intro_2}
\end{figure*}

Increasing integration -- i.e., decreasing the number of discrete hardware components used for quantum control will make qubit control more scalable. At present, there are two broad approaches for increasing integration: (1) Using off-the-shelf RFSoCs (2) Building custom Qubit Control ASICs. RFSoCs integrate an FPGA, a processor, DACs, and ADCs on the same chip and are typically used for telecommunication applications. RFSoCs can synthesize signals with tens of gigahertz bandwidth. Recently proposed ``QICK" and ``ICARUS-Q" platforms leverage RFSoCs to control qubits~\cite{qick,icarus}. Figure \ref{fig:fig1} shows how Qubit Control with RFSoCs would differ from a distributed approach. In theory, RFSocs are capable of supporting 100+ qubits per board (with the help of frequency-division multiplexing (FDM))~\cite{qick}. Compared to designing ASICs that can integrate a large number of DACs and ADCs on a chip, RFSoC based solutions offer flexibility that is crucial for near term qubit control as quantum hardware is rapidly evolving. As discussed in~\cite{qick}, using FDM with high-bandwidth DACs and ADCs available on these RFSoCs can increase the number of qubits controlled by one board, thus reducing the total cost and complexity of the setup. However, controlling more than thousands of qubits would require custom control hardware and new quantum-classical interfaces. We envision the following transition for qubit control as we scale the number of qubits.  
\begin{center}
\textsc{fpga} controllers $\xrightarrow{}$ \textsc{rfs}o\textsc{c} controllers $\xrightarrow{}$ \textsc{asic} controllers
\end{center}


To scale beyond thousands of qubits, we will need qubit controller that can operate at cryogenic temperatures~\cite{Tannu2017,digiQ,Bardin2019,Frank2022,VanDijk2020_}. For cryogenic control chips, the cooling capacity of the dilution refrigerator  results in the ``Power Wall" -- the average power dissipated by these chips should not exceed the rated cooling power of the dilution refrigerator. To this end, several architectural and design proposals envision scalable qubit control using ultra-low power CMOS and SFQ technologies~\cite{Frank2022,VanDijk2020_,Bardin2019,digiQ,McDermott2018,tannu2017cryogenic}.  



%


With increasing integration, potentially more qubits can be supported per control chip, reducing both complexity and cost. However, the memory bandwidth and capacity required to generate the pulses will scale linearly with the number of qubits as each qubit device requires a unique pulse shape (waveform) to enable high fidelity gates~\cite{optimal2, optimal3, optimal4}. Furthermore, for performing multi-qubit gates, most architectures use a coupler between qubits, and for reading the qubit, the readout resonators are used, and even these non-qubit components use device-specific pulses to enable high-fidelity operations. As a result, the number of unique pulses grows super-linearly with the increasing number of qubits.For example, we estimate, on IBM quantum computers, each qubit device requires about 18KB to store single-qubit, two-qubit, and readout pulses. To support more gates, we require additional memory capacity. We estimate that a hundred-qubit quantum computer would require up to 5MB of memory for pulse shapes of basic gates. Moreover, we will require significantly more memory capacity to support complex gates. For example, we estimate that a single Toffoli gate can require $>$20KB to store a pulse applied on three qubits~\cite{kim2022high}. 

Memory bandwidth is the primary bottleneck in scaling the number of qubits. For example, on IBM quantum computers, two DACs per qubit (with a sampling rate of 4.54 Gigasamples/sec each) are required to perform a single-qubit gate. We need a waveform memory to stream samples with more than 16 GB/s bandwidth to such DACs. Furthermore, the peak bandwidth required scales linearly with the number of qubits. To perform concurrent gates on a hundred-qubit machines, we need more than 2 TB/s. Current qubit control setups such as the one used for Google's Sycamore chip solve this problem using distributed waveform memory across 30+ FPGAs. However, distributed control does not scale practically with increasing number of qubits.

This paper focuses on reducing bandwidth and capacity overhead to scale qubit control. Waveforms are designed to have a tight frequency spectrum to limit crosstalk and leakage errors. As a result, waveforms are smooth and highly compressible. We evaluate the compressibility of waveforms used on three IBM machines and see that on average, we can reduce the capacity overhead by 8x using Discrete Cosine Transform (DCT) based compression. We use DCT based compression due to its high energy compaction properties. Moreover, DCT can be efficiently implemented in hardware. 
The waveform memory is loaded before the execution starts and is not updated during the execution; current compilers update waveform libraries only during calibration.

In this paper, we propose {\em {\sc compaqt}: Compressed Waveform Memory Architecture for Scalable Qubit Control}. {\sc compaqt} mitigates the waveform bandwidth bottleneck by compressing waveform libraries at compile time and storing compressed waveforms in the waveform memory. To execute a gate, {\sc compaqt} decompresses the waveforms and then streams it to the DAC as shown in Figure~\ref{fig:intro_2}(a). {\sc compaqt} boosts the memory bandwidth as the hardware decompression engine expands the memory contents, increasing the data supplied to the DAC per unit time, as shown in Figure~\ref{fig:intro_2}(b). This bandwidth gain directly translates to a gain in the number of concurrent operations that a controller can support, as shown in Figure~\ref{fig:intro_2}(c). {\sc compaqt} trades memory bandwidth with computational bandwidth. To enable fast and efficient decompression, we use a windowed DCT algorithm during compression which is decompressed with a corresponding inverse DCT (IDCT) engine in hardware. Windowing reduces hardware complexity, but it also reduces net compressibility. We pick window sizes that maximize the net bandwidth benefits. 
Fixed/floating-point multipliers in the IDCT algorithm are a major challenge in building a high-speed and efficient decompression engine. We use an integer IDCT algorithm~\cite{tran1999fast} that replaces multipliers with shifters and adders to significantly improve the performance and efficiency of {\sc compaqt}.


{\noindent The key contributions of this paper are summarized below:}
\begin{itemize}
    \item \textbf{Waveform memory bandwidth is a bottleneck}: We show how the bandwidth demand reduces the scalability of RFSoC based controllers (Section~\ref{sec:problem}).
    \item \textbf{Gate Pulse Waveforms are highly compressible}: We show that waveforms can be effectively compressed using the Discrete Cosine Transform without degrading the gate fidelity (Section~\ref{sec:compression}).
    \item \textbf{COMPAQT Microarchitecture}: We present a compressed waveform memory architecture that can increase the number of qubits supported by RFSoC based controllers by 5x (Section~\ref{sec:arch}).
\end{itemize}

\section{Background}
This section gives a brief description of how qubits are controlled, the organization of a typical control computer, and the challenge of scaling waveform memory.

\subsection{Manipulating Qubits via Gate Pulses}

\begin{figure}
    \centering
    \includegraphics[width=\linewidth]{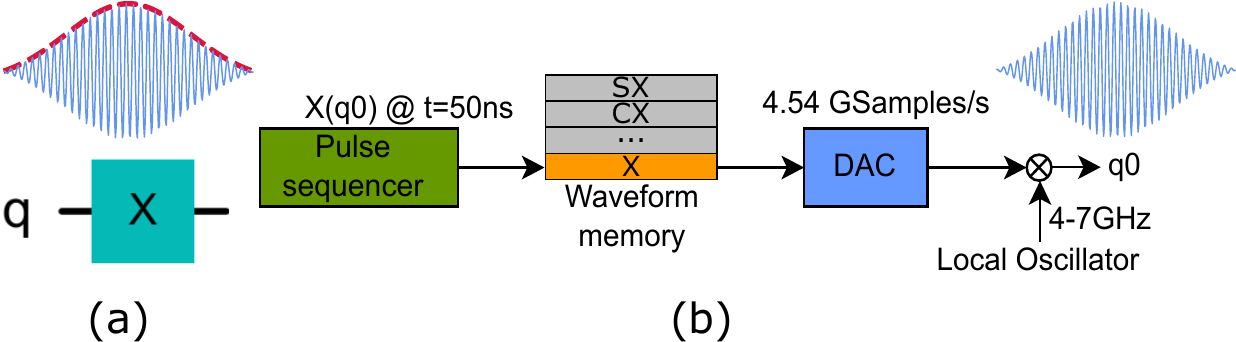}
    \caption{(a) An X-gate and its corresponding microwave pulse (in blue) The envelope of the pulse is shown in red (b) A typical pipeline for generating control pulses.
    }
    \label{fig:fig2}
\end{figure}

Quantum algorithms can be expressed as quantum circuits, a sequence of gate operations applied on qubit variables. On superconducting architectures, the gates are implemented using microwave pulses as qubit devices (such as transmons) are essentially oscillators that respond to signals in the frequency range of 4-10 GHz. Quantum control refers to the synthesis of these microwave pulses to implement various quantum gates. These microwave pulses can have different amplitudes, shapes, and duration depending on the type of gate and the qubit architecture. The state of a qubit is represented by a vector on a Bloch sphere~\cite{Nielsen2000-nt}. When a pulse is applied, the state rotates about the X/Y/Z axis of the Bloch sphere. Since Z rotations can be implemented in software~\cite{McKay2017}, microwave pulses are composed of two components that control the qubit state evolution about the X and Y-axis of the Bloch sphere. Consider the circuit shown in Figure~\ref{fig:fig2}(a), with X-gate applied on qubit \texttt{q1}, which inverts the qubit state. Physically, this gate is implemented as a pulse with a frequency equal to the qubit resonant frequency (shown in blue). The envelope of the pulse refers to its shape (shown as dotted line), and we will refer to it as a waveform henceforth. 

The physical implementation of two-qubit (2Q) gates such as \texttt{CNOT/CZ} depends on qubit architecture~\cite{Krantz2019}. For example, IBM machines use fixed-frequency qubits, wherein 2Q gates are implemented using Cross-Resonance (CR) gates~\cite{Rigetti2010, Chow2011} whereas the frequency-tunable qubits used by Google implement a 2Q gate called the \texttt{iSWAP} gate~\cite{Arute2019, Krantz2019}. Both implementations may require additional pulses for the coupler that enables interactions between neighboring qubits. In addition to standard gates, custom pulses can significantly boost application fidelity. For example, \texttt{Toffoli} gates have been demonstrated on superconducting hardware~\cite{kim2022high, Stojanovi2012}. Furthermore, Shi et al. proposed to use custom pulses to enable multi-qubit operators to reduce circuit depth and improve fidelity~\cite{Shi2019}, and Xie et al. showed that carefully re-shaping the waveforms can reduce ZZ-crosstalk~\cite{Xie2022}. 

\subsection{Control Computer Organization}


\begin{figure}
    \centering
    \includegraphics[width=\linewidth]{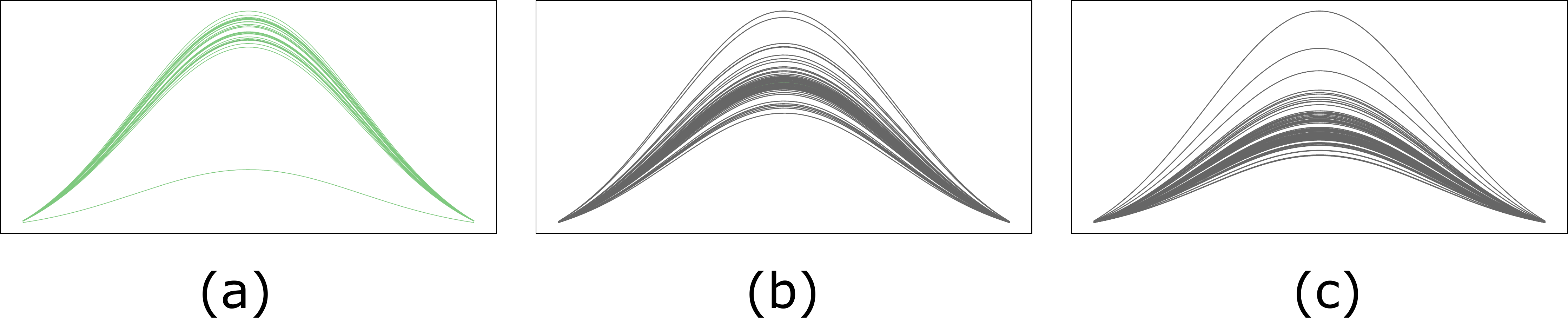}
    \caption{ $\pi$-pulse shapes of all (a) 27 qubits on IBM Toronto (b) 65 qubits on IBM Brooklyn (c) 127 qubits on IBM Washington machines.}
    \label{fig:pulse_diversity}
\end{figure}

Figure~\ref{fig:fig2}(b) shows a simplified control pipeline used to generate microwave gate pulses to manipulate the state of the qubit. A pulse sequencer is used to trigger the execution of gates (an X gate for qubit \texttt{q0} in this case). The pulse envelopes are stored in the waveform memory, and they consist of the Inphase (I) and Quadrature (Q) components for rotating the qubit about the X- and Y-axis of the Bloch sphere. Since DACs have a fixed sampling rate (sampling rate of IBM machines annotated in the figure), the waveform memory interface should supply data with a bandwidth that matches the sampling rate to ensure signal integrity. Higher DAC sampling rates are desirable for high-fidelity gates. The waveforms are converted to pulses at an intermediate frequency by a DAC. The resultant pulse is then mixed with a microwave carrier generated by a local oscillator (LO) to produce the final microwave pulse. 

State-of-the-art quantum control hardware uses discrete FPGAs,  DACs, and ADCs placed at room temperature. Controllers can be made more scalable with the use of RFSoC platforms that integrate a CPU, FPGA, DACs, and ADCs on the same chip. The CPU is interfaced with the control hardware to  calibrate qubits and load experiments, while the qubits are actively controlled using an FPGA. As suggested in \cite{qick}, a single RFSoC can control more than 100 qubits with the help of frequency division multiplexing (FDM), making this approach more scalable in terms of cost and complexity. The high-bandwidth DACs/ADCs on-chip eliminate the need for external analog mixers; pulses can be directly synthesized at the desired frequency. Other ways of implementing the control computer include ASICs that can integrate large numbers of discrete components but are expensive to design and manufacture.

\subsection{Challenge: Scaling Waveform Memory}

Unfortunately,  every qubit has a unique waveform for every physical gate that the system supports. As shown in Figure \ref{fig:pulse_diversity}, every qubit on IBM quantum computers has a different $\pi$ pulse, which is tuned to maximize gate fidelity. These waveforms are periodically updated via calibration cycles. Furthermore, the connections between adjacent qubits (couplers) are unique, and this results in different waveforms used for controlling the qubit and the coupler for multi-qubit gates like the \texttt{CNOT} or \texttt{Toffoli}. Finally, pulses used for readout have waveforms that depend on the qubit. 

With the increasing number of qubits, the total number of unique waveforms that have to be stored in the waveform memory increase linearly. Furthermore, this increase in capacity requirements will be accompanied by an increase in memory bandwidth requirements, since more qubits would need to be driven concurrently, especially for running quantum error correction circuits. While this increase in memory capacity and bandwidth requirements is not a bottleneck with brute-force scaling of control, the next section will show how the memory architecture in integrated RFSoC controllers can limit scalability.




\begin{sloppypar}
\section{Waveform Memory Bottleneck}
\label{sec:problem}
\end{sloppypar}

In this section, we will show the impact of linearly scaling the bandwidth demand of qubit control on FPGA-based systems. We use parameters described in the Table~\ref{tab:params} to model capacity and bandwidth requirements, which are derived from IBM~\cite{ibm_systems, Alexander2020} and Google~\cite{Arute2019, weber, McEwen2021} systems.  

\begin{table}[hpt]
\caption{Parameters used to estimate memory capacity and bandwidth for controlling one qubit.}
\label{tab:params}
\begin{center}
\begin{small}
\setlength{\tabcolsep}{0.05cm} 
\renewcommand{\arraystretch}{1.2}
\scalebox{0.71}{
\begin{tabular}{|c|c|c|c|c|c|c|}
\hline
\multicolumn{1}{|l|}{\textbf{Vendor}} & \textbf{\begin{tabular}[c]{@{}c@{}}Sampling\\ Rate($f_s$)\end{tabular}} & \textbf{\begin{tabular}[c]{@{}c@{}}Sample\\ Size($N_s$)\end{tabular}} & \textbf{Gate Set}                                                & \textbf{\begin{tabular}[c]{@{}c@{}}Gate\\ Latencies($\tau$)\end{tabular}}          & \textbf{Connectivity}                                     & \textbf{\begin{tabular}[c]{@{}c@{}}Memory\\ (per qubit)\end{tabular}} \\ \hline
IBM                                   & 4.54GS/s                                                           & 32-bits                                                          & X, SX, CX                                                        & \begin{tabular}[c]{@{}c@{}}1Q:30ns;\\ 2Q:300ns;\\ Readout:300ns\end{tabular} & \begin{tabular}[c]{@{}c@{}}Heavy\\ hexagonal\end{tabular} & 18KB                                                                  \\ \hline
Google                                & 1GS/s                                                              & 28-bits                                                          & \begin{tabular}[c]{@{}c@{}}fsim, iSWAP,\\ phased XZ\end{tabular} & \begin{tabular}[c]{@{}c@{}}1Q:25ns;\\ 2Q:30ns;\\ Readout:500ns\end{tabular}  & Grid                                                      & 3KB                                                                   \\ \hline
\end{tabular}
}
\end{small}
\end{center}
\end{table}


We estimate memory capacity (MC) and bandwidth (BW) required for controlling one qubit -
\begin{align*}
    MC = \sum^{n_{sq}}_{i=0} f_s.N_s.\tau_{i} +  \sum^{d*n_{tq}}_{j=0} f_s.N_s.\tau_{j} + f_s.N_s.\tau_{readout}
\end{align*}

The product of sampling frequency of DAC ($f_s$), sample size ($N_s$), and gate latency($\tau$) is the memory required for storing one gate waveform. We can perform a of total $n_{sq}$ types of single qubit gates and $n_{tq}$ types of two qubit gates. Furthermore, two qubit waveforms are unique for every pair of qubits, and thus for a qubit with $d$ neighboring qubits, we will need $d*n_{tq}$ waveforms. In addition to gates, each qubit will need a readout pulse with latency $\tau_{readout}$. As shown in Table~\ref{tab:params}, IBM uses two single-qubit gates and one type of two-qubit gate, which require a total of 18KB of waveform memory for every qubit device. Note that the sample size $s$ accounts for both I and Q channels. The required memory bandwidth of waveform buffer is determined by the sampling frequency and bit-width of DAC as shown below.  
\begin{align*}
    BW = f_s . s
\end{align*}



\begin{sloppypar}
\subsection{Scaling Capacity and Bandwidth Demand}
\end{sloppypar}

\begin{figure*}[t]
    \centering
    \includegraphics[width=0.95\textwidth]{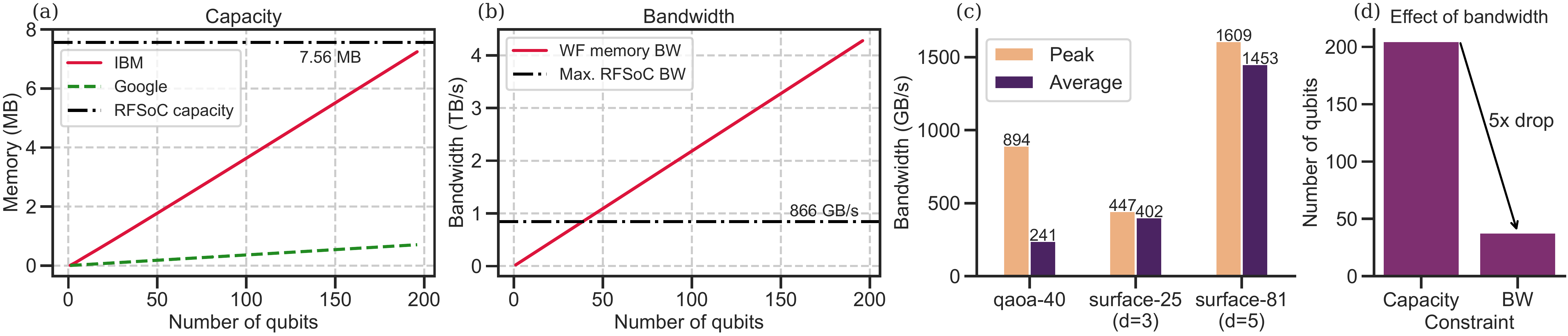}
    \caption{Waveform (WF) memory (a) capacity and (b) bandwidth (BW) requirements (c) Peak and average bandwidth for select benchmarks (d) For an RFSoC platform, if only capacity is considered a constraint, the number of qubits that can be supported is $>$200. If bandwidth is taken as a constraint, this number drops by 5x to less than 40.
    }
    \label{fig:mem_capacity}
\end{figure*}

\vspace{0.05in}
\noindent{\textbf{Capacity scaling.}}
Every qubit device requires a unique waveform, which is determined during the calibration process. It is essential to build control engine that supports the device specific waveforms to enable high gate fidelity. As a result, required memory capacity scales linearly with number of qubits. In addition to storing waveforms for qubits themselves, some architectures require waveforms for couplers that connect qubits to improve fidelity of two qubit (2Q) gates. The capacity required by the coupler waveforms depends on the connectivity, and this must be considered when determining the total capacity. Coupler waveforms are unique for every pair of qubits. As shown in Figure \ref{fig:mem_capacity}(a), the required waveform memory capacity scales linearly with the number of qubits in the system (some approximations made to account for coupler waveforms). 
The RFSoC capacity includes both Block-RAM (BRAM) and UltraRAM (URAM) blocks~\cite{xilinx} and has been added as a reference for understanding how the capacity requirements scale (both BRAMs and URAMs will be collectively referred as BRAMs from here on). 

\vspace{0.05in}
\noindent{\textbf{Bandwidth scaling.}}
Figure \ref{fig:mem_capacity}(b) shows the linear scaling of the required memory bandwidth to drive all qubits concurrently with the 6GS/s DACs found on RFSoCs. Using such high-bandwidth DACs eliminates the need for external analog mixing, requiring frequent re-calibration. Google/IBM systems will show a similar, albeit lesser bandwidth scaling since those systems employ brute-force scaling with discrete DACs and FPGAs. The reference RFSoC bandwidth\footnote{Determined by assuming a 16x difference in clock speeds of the FPGA and DAC and considering 1260 BRAMs~\cite{xilinx}.} shows that a single RFSoC can concurrently control less than 40 qubits, as FPGA clocks are significantly slower than DAC sampling rates. For example, On the QICK platform, the DAC is 16x slower compared to waveform memory.

\vspace{0.05in}

\noindent{\textbf{Circuit Scalability.}} To understand the impact of waveform memory bandwidth, we estimate the peak and average bandwidth required for representative circuits that users would want to run in the near future. Average bandwidth is determined by how many gates are performed concurrently on an average, whereas peak bandwidth is determined by the maximum concurrency needed for executing the circuit. Shown in Figure \ref{fig:mem_capacity}(c), which plots the peak and average bandwidth for three benchmarks -- Quantum Approximate Optimization Algorithm (QAOA) with 40 qubits, a distance-3 surface code with 25 qubits, and a distance-5 surface code with 81 qubits. For surface codes, the difference between peak and average bandwidth is small as quantum error correction codes such as the surface code~\cite{Fowler2012} are designed to run concurrent gates as any delay in generating and measuring quantum error syndromes can significantly degrade the protection offered by the error correction code. NISQ applications such as QAOA are not bandwidth intensive on average. However, the last step of all NISQ circuits involves the concurrent measurement of all qubits, requiring the maximum possible waveform memory bandwidth. Unfortunately, serializing measurements is not an option as it can significantly degrade readout fidelity.

\vspace{0.05in}

\begin{center}
\emph{In summary, for RFSoCs, bandwidth becomes a bottleneck as they can support the capacity but not the bandwidth to run surface code circuits with highly concurrent gates.} 
\end{center}

\vspace{0.02in}

\subsection{Meeting Capacity and Bandwidth Demand}





The demand for capacity and bandwidth of waveform memory scales linearly with the number of qubits. This presents a challenge for scaling quantum control computers, irrespective of whether it is implemented on an RFSoCs or an ASIC. On RFSoCs, the internal memory bandwidth and on-chip memory capacity can limit how many qubits we can control. We can boost the memory bandwidth by interleaving (or banking) the memory but due to rigid constraints on Block RAMs, the peak internal bandwidth is limited.  

On cryogenic ASICs, power is the primary constraint and thus providing the necessary capacity and bandwidth within a limited power budget is challenging. Furthermore,  techniques to improve scalability of qubit control, such as Frequency Division Multiplexing (FDM), cannot be supported if the waveform memory is too small or if it cannot provide the necessary bandwidth. For example, the QICK framework can potentially control 100+ qubits per RFSoC board with FDM. However, before waveforms are mixed onto a single channel for FDM, the waveforms for all the multiplexed qubits must be stored and then individually generated, which means that the waveform memory must have sufficient capacity and bandwidth for all qubits. As shown in Figure~\ref{fig:mem_capacity}(d), the bandwidth on RFSoC platforms is insufficient to control more than 40 qubits.

\section{Waveform compression}
\label{sec:compression}

In this section, first we give an overview of {\sc compaqt}, and then discuss different methods that can be used to enable compression and their impact on the gate fidelity. 

\subsection{COMPAQT: Insight and Overview}

\begin{figure} [t]
    \centering
    \includegraphics[width=0.8\linewidth]{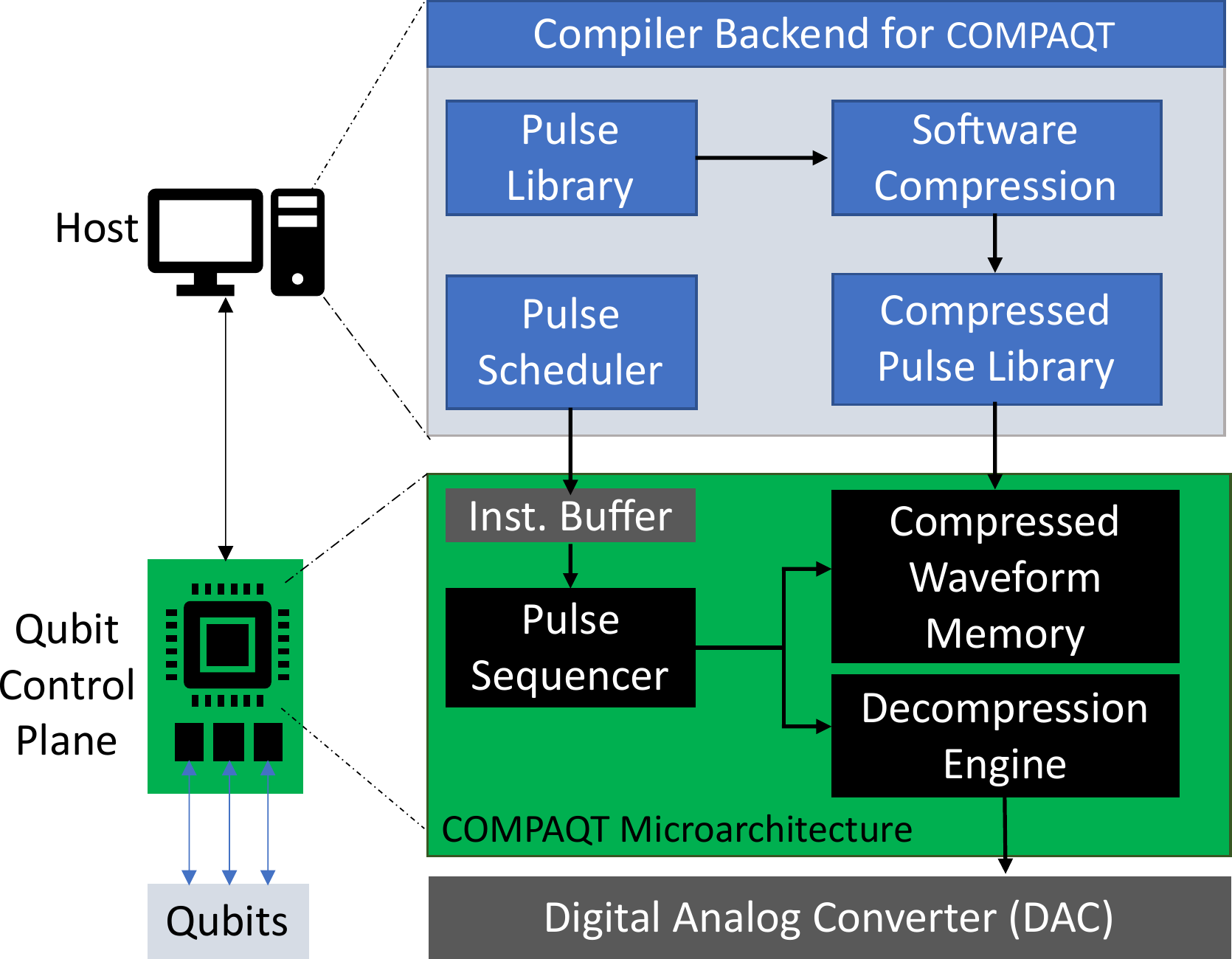}
    \caption{Architectural overview of COMPAQT}
    \label{fig:overview}
\end{figure}

{\noindent \textbf{Insight.}} We observe that  waveforms are highly compressible as they are designed to have a tight bandwidth and low spectral leakage, which means they are smooth and change slowly in time-domain.  We leverage this property to alleviate the high waveform memory bandwidth. We propose {\em {\sc compaqt}: Compressed Waveform Memory Architecture for Scalable Qubit Control}. By compressing the waveforms, we reduce the required capacity, and by using a low-latency decompression engine, we can quickly expand the compressed waveform to meet the desired memory bandwidth demand. 

{\noindent \textbf{Design Overview.}} Figure~\ref{fig:overview} shows overview of {\sc compaqt}, where waveforms are compressed in the software and then transferred to the controller waveform memory. The pulse sequencer can then play waveforms by decompressing each sample and then passing it to the DAC. {\sc compaqt} leverages the read-only nature of the Waveform Memory, which is loaded with waveforms at the end of a calibration cycle. To reduce capacity and bandwidth overhead, {\sc compaqt} compresses the waveforms such that only a few samples are needed for representing the entire waveform. 
Since waveform memory is a ROM, compressing the waveforms has no hardware overhead and can be done in purely software while decompressing the waveforms must be done entirely on hardware. Note that there could be other (and better) compression techniques, the techniques used in this paper are purely to demonstrate the compressibility of waveforms and the advantages compression offers for improving scalability. 

\begin{figure*}[t]
    \centering
    \includegraphics[width=0.9\linewidth]{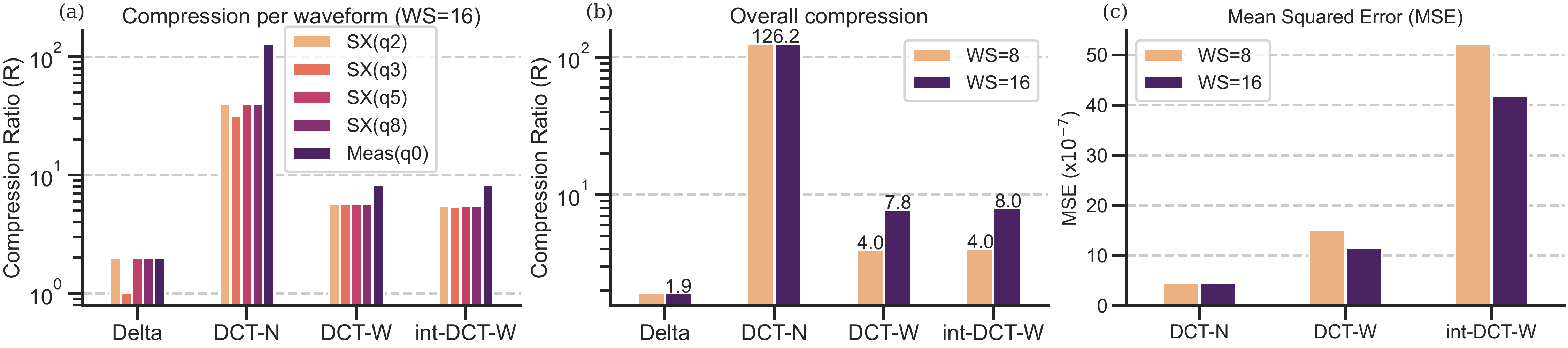}
    \caption{(a) Compression Ratio (old size / new size) for five waveforms on IBM Guadalupe (b) Overall Compression Ratio for qft-4 (c) Average Mean Squared Error (MSE) for all waveforms of qft-4. }
    \label{fig:compression}
\end{figure*}


\subsection{Selecting the Compression Algorithm}

\textbf{Base-delta and dictionary-based compression} are widely studied algorithms to maximize on-chip memory capacity~\cite{pekhimenko2012base,alameldeen2004adaptive}. However, traditional memory compression algorithms are not effective in compressing waveform data. Delta compression reduces the number of bits by storing only the difference between adjacent samples. However, waveforms can have zero crossings or sign changes. If samples are represented by signed data types (fixed or floating point), the zero crossings will result in a difference that occupies the entire bit-field of the original samples. Furthermore, any non-uniform deltas will add complexity to decompression. Our evaluations on IBM hardware, show that the size of waveforms with no zero-crossings can be reduced by up to 2x as deltas generally occupy half the original bit-width of the sample. However, there is no reduction for waveforms with one or more zero crossings, as shown in Figure~\ref{fig:compression}(a). Moreover, in base-delta compression there is a sequential dependence between samples, which limits its utility for improving bandwidth. Similarly, other widely used dictionary-based compression schemes are not effective, as waveform sample values can have arbitrary values, which rarely repeat.    

We propose to use \textbf{transform-based compression}, which is widely used for audio and video codecs. These compression schemes leverage the Discrete Cosine Transform (DCT) that transforms the input data from the spatial domain to the frequency domain~\cite{ahmed1974discrete}, where data can be represented with significantly less memory~\cite{recommendation199281, sze2014high}. Specifically, the DCT of the input signal results in high-energy components in the first few samples after which the magnitude of the transformed samples becomes significantly smaller. The original signal can be recovered by computing the inverse (IDCT) of the transformed sequence.

\begin{figure}[b]
    \centering
    \includegraphics[width=0.9\linewidth]{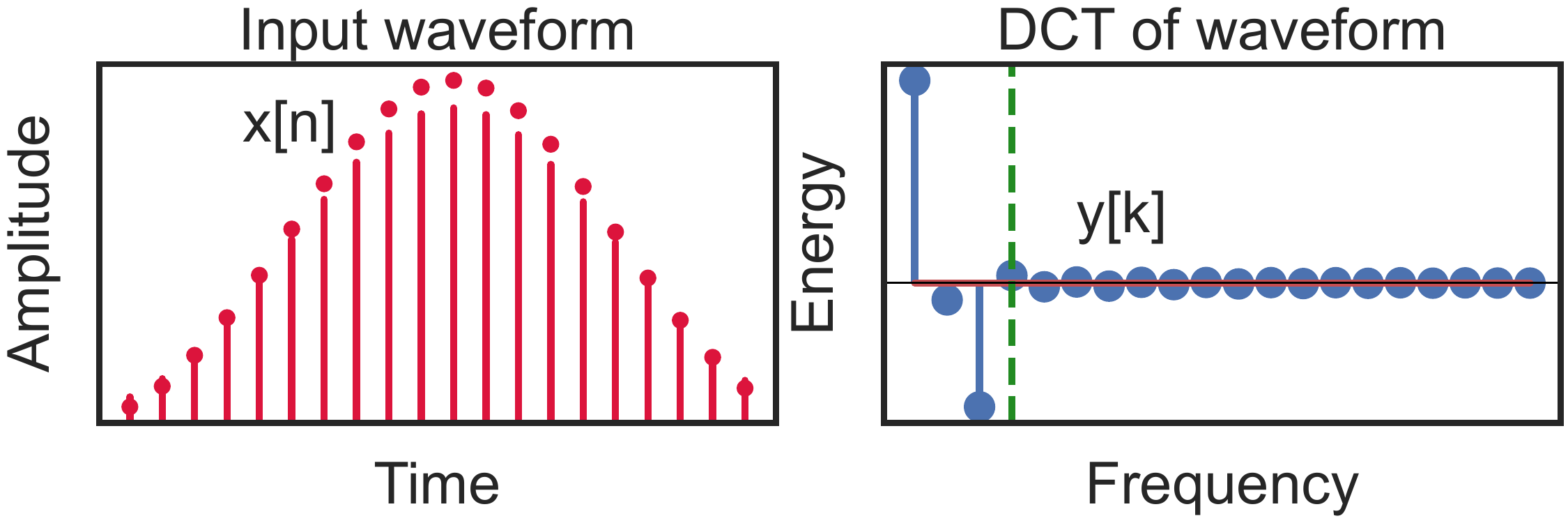}
    \caption{Input waveform and its DCT, the vertical green line shows the start of RLE where $|$sample$|$ $<$ threshold.}
    \label{fig:dct}
\end{figure}






\subsection{DCT-based Waveform Compression}
\label{sec:dct_compression}



{\noindent \textbf{Discrete Cosine Transform (DCT).}} The N-point DCT of a signal $x[n]$ and its inverse (IDCT) can be computed using Equations~\ref{eqn:dct} and \ref{eqn:idct} respectively. 

\begin{small}
\begin{equation}
    \label{eqn:dct}
    y[k] = {\frac{1}{\sqrt{N}}}\sum_{n = 0}^{N - 1} x[n]cos\left[ \frac{\pi(2n + 1)k}{2N}\right] 
\end{equation}
\end{small}

\begin{small}
\begin{equation}
    \label{eqn:idct}
    x[k] = \frac{y[0]}{\sqrt{N}} + \sqrt{\frac{2}{N}}\sum_{n = 1}^{N - 1} y[n]cos\left[ \frac{\pi(2k + 1)n}{2N}\right] 
\end{equation}
\end{small}

In \textsc{compaqt}, decompression is performed on hardware, and since it is part of the critical path during waveform synthesis, it must have low latency and a minimal hardware footprint. If a waveform with $N$ samples is compressed with an N-point DCT, decompressing it would be complex as 1) $N$ can vary for different waveforms, and 2) $N$ can be very large -- a two-qubit CR waveform on IBM systems can have 1000+ samples.
For this reason, we use a windowed DCT scheme in which the input waveform is broken into windows of a fixed size called a {\em window size (WS)}. The windowed DCT scheme shall hereon be abbreviated as \texttt{DCT-W} while the N-point scheme shall be abbreviated as \texttt{DCT-N}.

The compression algorithm using the DCT is as follows: The DCT of a window of the input waveform is computed (for \texttt{DCT-N}, the window would be the entire waveform) after which a thresholding operation is performed, which sets samples with low magnitudes equal to zero. Finally, Run-Length Encoding (RLE)~\cite{bradley1969optimizing} is performed on the transformed window. RLE replaces all the zeros with a single codeword that contains two fields: 1) a signature identifying it as the RLE codeword and 2) the number of zeros that have been encoded to yield the compressed waveform.


\begin{table}[hpt]
\caption{Implemented DCT variants.}
\vspace{-0.2in}
\label{tab:dct_variants}
\begin{center}
\begin{normalsize}
\setlength{\tabcolsep}{0.05cm} 
\renewcommand{\arraystretch}{1.2}
\scalebox{0.73}{
\begin{tabular}{ | c | c | c | }
\hline
    \textbf{Variant}    &           \textbf{Description}                 & \textbf{Hardware Complexity}  \\ \hline \hline
    
    \texttt{DCT-N}      &           N-point DCT, WS = length of waveform & High  \\ \hline
    \texttt{DCT-W}      &           Windowed DCT, WS = {8, 16}           & Moderate   \\ \hline
    \texttt{int-DCT-W}  &           Windowed integer-DCT, WS = {8, 16}   & Low   \\ \hline
    
\end{tabular}
}
\end{normalsize}
\end{center}
\end{table}

The compression using \texttt{DCT-N} has been visualized in Figure~\ref{fig:dct} for single-qubit gates. We use typical Derivative Removal by Adiabatic Gate (DRAG) waveforms used in most superconducting systems to implement single qubit gates. RLE is started only when the transformed waveform after thresholding is consistently zero. This process is done separately for I and Q channels, and to simplify the decompression, the number of samples per window after compression are kept the same for both channels.

\vspace{0.05in}
{\noindent \textbf{Integer-DCT.}}
The DCT/IDCT in its original form requires a minimum number of fixed/floating-point multipliers that make hardware implementations very resource intensive. The 8-point DCT/IDCT (WS=8) requires a minimum of 11 multipliers~\cite{loeffler1989practical}. Larger transforms need resources, with the 16-point DCT/IDCT requiring a minimum of 26 multipliers.


To make waveform compression using DCT scalable for quantum control applications, we also implemented an integer DCT consistent with the HEVC standard~\cite{sze2014high}. Since this variant of the DCT was aimed at making hardware implementations as efficient as possible, we do not use the arbitrary N-point transform. The windowed integer DCT will henceforth be abbreviated as \texttt{int-DCT-W}. Since the DCT coefficients are scaled by a constant factor, the input waveform is scaled by the same factor in software (this can be done in hardware too). The constant scaling factor $S$ for an N-point DCT is given by $S = 2^{6 + \frac{log_2N}2}$.
All DCT variants are summarized in Table \ref{tab:dct_variants}. We used two different window sizes to evaluate the compression of waveforms using DCT. 

\vspace{0.05in}

{\noindent \textbf{Fidelity-Aware Thresholding.}}
Each gate pulse is unique and a uniform threshold during compression can cause fidelity loss for some qubits. To enable high fidelity gates, we find optimal threshold for every gate pulse. We use the Algorithm~\ref{alg:COMPAQT}, to tune the threshold to obtain the target gate fidelity. We observe that MSE between decompressed and uncompressed pulses are highly co-related to the gate fidelity, and we can use MSE for tuning the threshold at compile time. We can take a step further and integrate the Fidelity-Aware compression within the gate calibration loop.

{
\setlength{\textfloatsep}{0pt}
\SetAlgoNoLine
{
\begin{algorithm}[htb]
\caption{ Fidelity-Aware Compression}
\label{alg:COMPAQT}
\SetKwInput{KwInput}{Input}                
\SetKwInput{KwOutput}{Output}              
\DontPrintSemicolon
    \KwInput{ $W_{in}$ (Gate Pulse), $\varepsilon$ (Target Error) }\vspace{0.0 in}
    \KwOutput{ $W_{out}$ (Compressed Pulse)}\vspace{0.05 in}
  \SetKwFunction{compaqt}{Get\_Compressed\_Pulse}
  \SetKwProg{Fn}{Function}{:}{}
  \Fn{\compaqt{$W_{in}$, $\varepsilon$}}
  {
  \While{mse $\leq$ ~$\varepsilon$}{
      Y = int-{\sc DCT}($W_{in}$, $W_{out}$) \CommentSty{\color{red} // Compute DCT} \;
      \For{s \textbf{in} Y} 
        {
        \lIf{s $<$ threshold}{s = 0}
        }
    $W_{out}$ = int-i{\sc dct (Y)}\;
    mse = get\_{\sc mse} ($W_{in}$, $W_{out}$)\;
    threshold = threshold/2\;
    \lIf{ $threshold <10^{-6}$}{\Return ~-1 }\CommentSty{\color{red} // No solution found}\;
    } 
  }
  \KwRet{$W_{out}$}
\end{algorithm}}
}






\subsection{Efficacy of DCT Compression}
The DCT-based compression scheme is lossy and its
effectiveness depends on two factors: 1) Compressibility of the waveforms. 2) Fidelity of decompressed waveforms. To evaluate compressibility, we measure the reduction in memory size to store all waveforms used for a benchmark circuit. For evaluating the effect on fidelity, we ran two-qubit Randomized Benchmarking~\cite{Magesan_2011} on various IBM machines. 

\vspace{0.05in}
{\noindent \textbf{Compressibility.}}
Figure \ref{fig:compression}(a) shows the reduction ($R =old\ size / new\ size$) in required memory for five representative waveforms of \texttt{qft-4} benchmark derived from~\cite{li2021qasmbench} implemented on the IBM Guadalupe machine. 
Figure \ref{fig:compression}(b) shows the overall reduction in memory for \texttt{qft-4}. Note that measurement and 2Q gates are longer and more compressible than 1Q gates which affects the total compression ratio for \texttt{qft-4}. A window size of 8 has the least reduction because RLE is limited to a maximum of 8 samples at a time. Increasing the window size makes the IDCT design complex. Thus, more compression can be achieved at the cost of hardware complexity. Delta compression shows an overall reduction in size for the \texttt{qft-4} circuit but not all waveforms are compressible due to the overhead of supporting zero crossing as shown in Figure \ref{fig:compression}(a). 



\begin{figure} [t]
    \centering
    \includegraphics[width=\linewidth]{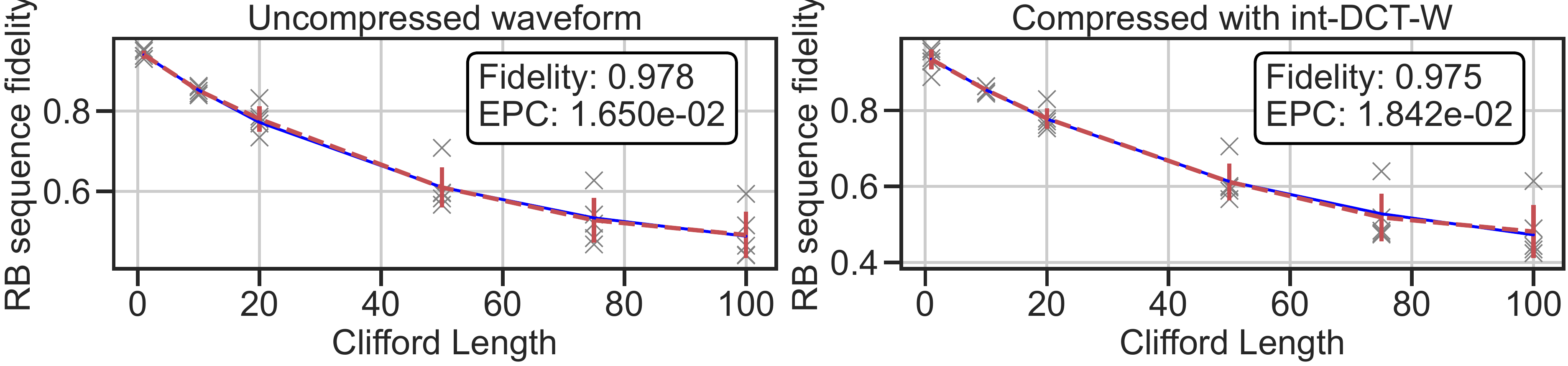}
    \caption{Randomized Benchmarking of two qubit system with the uncompressed baseline gate pulses and compressed gate pulses with int-DCT-W on IBM Guadalupe.}
    \label{fig:RB}
\end{figure}

\vspace{0.05in}
{\noindent \textbf{Impact on gate fidelity.}}
Distortion in the waveforms can reduce gate fidelity. To understand whether compression distorts the waveforms, we compute Mean Squared error (MSE) between the original waveforms and the decompressed waveforms. Figure~\ref{fig:compression}(c) shows the average MSE over all waveforms used for the \texttt{qft\_n4} benchmark for three variants of DCT-based compression. The MSE lies between $10^{-5}$ and $10^{-7}$, and it is highest (but still negligible) for the \texttt{int-DCT-W} variant since it incorporates integer approximations before the thresholding step in the algorithm.

Moreover, to test the gate fidelity, first we compress and then decompress waveforms by emulating the decompression engine design, and then we use these decompressed waveforms to drive qubits on IBM quantum computers. Figure~\ref{fig:RB} shows the RB fidelity with increasing number of Clifford gates and average Error Per Clifford (EPC) which averages the total error over the number of gates. Table~\ref{tab:tomo} summarizes the fidelity of two-qubit Randomized Benchmarking (RB) on three IBM machines. The results show that even though DCT based compression is lossy, the decompressed waveform is similar to the original, because of which there is minimal effect on the fidelity of the gates. 
\begin{table}[h]
\caption{RB fidelity for uncompressed baseline pulse and compressed pulses with window size of 16.}
\label{tab:tomo}
\begin{center}
\begin{footnotesize}

\scalebox{0.9}{
\begin{tabular}{|c|c|c|c|}
\hline
\textbf{Design} & \textbf{\begin{tabular}[c]{@{}c@{}}2Q RB Fidelity\\ IBM Bogota\end{tabular}} & \textbf{\begin{tabular}[c]{@{}c@{}}2Q RB Fidelity\\ IBM Guadalupe\end{tabular}} & \textbf{\begin{tabular}[c]{@{}c@{}}2Q RB Fidelity\\ IBM Hanoi\end{tabular}} \\ \hline
Baseline            & 0.980                                                                     & 0.978                                                                           & 0.987                                                                       \\ \hline
\texttt{DCT-N}      & 0.982                                                                     & 0.977                                                                           & 0.989                                                                       \\ \hline
\texttt{DCT-W}      & 0.983                                                                     & 0.976                                                                           & 0.986                                                                       \\ \hline
\texttt{int-DCT-W}  & 0.983                                                                     & 0.975                                                                           & 0.988                                                                       \\ \hline
\end{tabular}
}
\end{footnotesize}

\end{center}
\end{table}








\section{COMPAQT Microarchitecture}
\label{sec:arch}
In this section, we discuss the architecture of the decompression pipeline, the decompression engine, and the waveform memory. Furthermore, we will discuss adaptive decompression strategies to make {\sc compaqt} scalable with cryogenic ASICs.  


\subsection{Decompression Pipeline Architecture}





\begin{figure}[t] 
    \centering
    \includegraphics[width=0.85\linewidth]{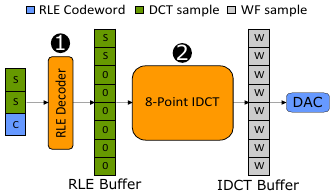}
    \caption{Decompression pipeline for waveforms compressed with DCT-W/int-DCT-W (WS=8).}
    \label{fig:decompression}
\end{figure}

\begin{figure}[b]
    \centering
    \includegraphics[width=0.75\linewidth]{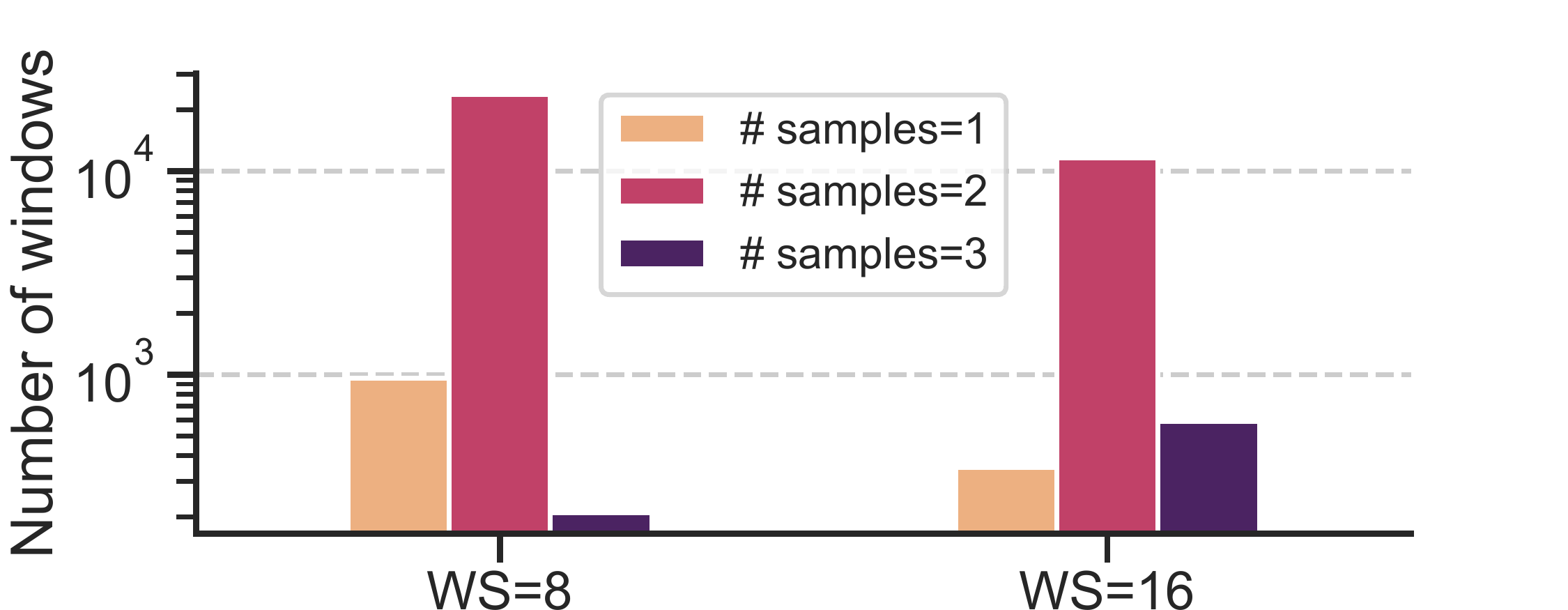}
    \caption{Histogram of the number of samples per window in the compressed waveform (including the RLE codeword) for 132 waveforms from IBM Guadalupe.}
    \label{fig:max_samples}
\end{figure}

\begin{figure*}
    \centering
    \includegraphics[width=0.95\linewidth]{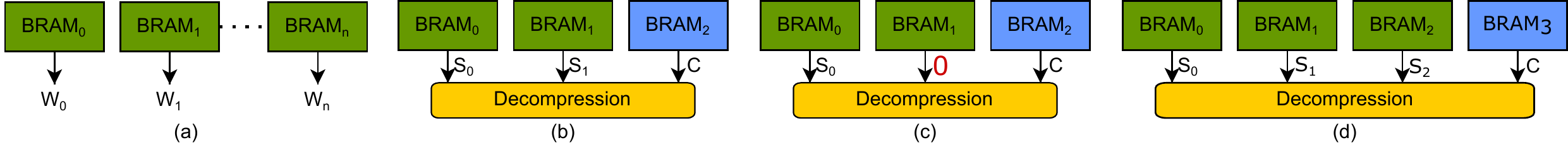}
    \caption{(a) Interleaved waveform samples in different BRAMs to boost bandwidth (b) Reduction in number of BRAMs for a waveform with maximum of three samples per window (including RLE code word) (c) Compressed window with two non-zero samples interleaved in three BRAMs (d) Interleaved memory organization for a pulse that requires four samples for a worst-case window.}
    \label{fig:fpga_mem}
\end{figure*}

Decompression is a two-step process, as shown in Figure~\ref{fig:decompression}. In \circled{1}, the codeword generated by RLE is decoded to yield all the encoded zeros and then in \circled{2}, the IDCT of the resultant vector produces the decompressed waveform. To enable hardware efficient design, we use the windowed IDCT algorithm in {\sc compaqt}. If compression was done using a window size of sixteen, the samples generated by the RLE decoder and IDCT would scale accordingly. 

Compression reduces the waveform capacity, but, more importantly, by storing a small number of samples and expanding them in the RLE buffer, DCT based compression enables higher bandwidth. As shown in Figure~\ref{fig:decompression}, the three compressed waveform samples yield eight waveform samples after decompression, thereby boosting the waveform memory bandwidth. In our design, for simplicity and efficient hardware synthesis, we design compressed waveform memory with uniform width, which is determined by the size of the worst-case compressed window. For example, the width of the compressed DCT samples (green+blue squares) in Figure~\ref{fig:decompression} is three samples, including the RLE codeword. Figure~\ref{fig:max_samples} shows the histogram of the number of samples in every window for \texttt{int-DCT-W} showing that waveforms are highly compressible. Based on an empirical analysis, the worst-case number of samples the design must cater to is thus three as shown by the histogram. The design of the decompression pipeline will be complex if we enable each window to have a variable number of samples. In {\sc compaqt}, we choose a uniform input buffer size which sacrifices compressibility to enable a significant performance boost.



\subsection{Decompression Engine Design}

In the DCT decompression pipeline, the RLE decoder depends on the RLE codeword to decompress all zeros and feed them to the IDCT stage. The signature in the RLE codeword can be used to identify the codeword, and if the RLE codeword specifies that $c_n$ zeros have been encoded, the last $c_n$ inputs of the IDCT stage can be set to 0.





A major advantage of using \texttt{int-DCT-W} over \texttt{DCT-W} is that the multipliers used for computing the inverse transform can be replaced with shift-and-add operations, drastically reducing hardware complexity and improving performance. We encourage readers to peruse existing designs for \texttt{DCT-W}~\cite{bukhari2002dct, fang2002lightweight, shen2012unified} and \texttt{int-DCT-W}~\cite{zhou2022effective, singhadia2020hardware, chatterjee2018optimized} (IDCT circuits are simply the reverse of DCT circuits).


\begin{table}[hpt]
\caption{Hardware resources needed for the IDCT engine of DCT-W and int-DCT-W.}
\label{tab:idct_hw}
\begin{center}
\begin{small}
\setlength{\tabcolsep}{0.05cm} 
\renewcommand{\arraystretch}{1.2}
\scalebox{0.95}{
\begin{tabular}{|c|c|c|c|c|}
\hline
\textbf{Variant}         & \textbf{Window size(WS)} & \textbf{Multipliers} & \textbf{Adders} & \textbf{Shifters} \\ \hline \hline
\texttt{DCT-W}           & \multirow{2}{*}{8}   & 11                   & 29              & 0                 \\ \cline{1-1} \cline{3-5} 
\texttt{int-DCT-W}       &                      & 0                    & 50              & 26                \\ \hline
\texttt{DCT-W}           & \multirow{2}{*}{16}  & 26                   & 81              & 0                 \\ \cline{1-1} \cline{3-5} 
\texttt{int-DCT-W}       &                      & 0                    & 186             & 128               \\ \hline
\end{tabular}
}
\end{small}
\end{center}
\end{table}

Table \ref{tab:idct_hw} summarizes the differences in requirements for implementing the IDCT in hardware for both \texttt{DCT-W} and \texttt{int-DCT-W} with a window size of eight. The \texttt{DCT-W} design is based on Loeffler's algorithm and is the least expensive implementation~\cite{loeffler1989practical}. For \texttt{int-DCT-W}, the multiplications are converted to shift-and-add operations~\cite{singhadia2020hardware}. This optimization enables the IDCT engine to have a constant latency of one clock cycle.


\subsection{Banked Compressed Waveform Memory}



Typically, the RFSoC FPGA clock is slower ($\approx$0.3~GHz) than the DAC sampling frequency ($\approx$6~GHz). As a result, even in the baseline uncompressed systems, feeding data to the DACs at the peak sampling rate can become challenging. To solve this problem, waveform samples for a single waveform are interleaved in multiple BRAMs. By banking waveform memory, we can compensate for the difference in frequencies between the FPGA and the DAC as shown in Figure \ref{fig:fpga_mem}(a). However, this approach is not scalable and eventually hits a BRAM bandwidth wall due to the finite number of BRAMs available. In {\sc compaqt}, we use memory interleaving with compressed waveforms, as fewer BRAMs are needed for a waveform with the decompression pipeline. 


Figure \ref{fig:max_samples} showed that regardless of the window size, the number of samples in the compressed waveform is $\leq$ 3 for \texttt{int-DCT-W}. Figure \ref{fig:fpga_mem}(b) is an example showing how if the waveform samples were originally interleaved in multiple BRAMs, the number of BRAMs needed after compression with WS=16 would be just three (if WS=8 was used, the number of BRAMs would have been six, since two 8-point IDCT modules would be needed). The decompression block in the figure is the pipeline shown in Figure \ref{fig:decompression} and the results after decompression can be stored on distributed memory in the RFSoC until they are used. Figure~\ref{fig:fpga_mem}(c) shows how the input to the decompression engine is zero if the number of samples in the window is less than the maximum number and Figure~\ref{fig:fpga_mem}(d) show a case where the maximum number of samples per window is four.

\begin{table}[hpt]
\caption{Number of qubits that can be supported by an FPGA based design with uncompressed waveform memory and a design with compressed memory using int-DCT-W for different window sizes.}
\label{tab:gain_qubits}
\begin{center}
\setlength{\tabcolsep}{0.25 cm} 
\renewcommand{\arraystretch}{1.1}
\scalebox{0.90}{
\begin{tabular}{|c|c|c|c|}
\hline
\textbf{}                                                                        & \textbf{Uncompressed} & \textbf{WS=8} & \textbf{WS=16} \\ \hline 
\textbf{\begin{tabular}[c]{@{}c@{}}Number of qubits\\ (normalized)\end{tabular}} & 1                     & 2.66          & 5.33           \\ \hline
\end{tabular}
}
\end{center}
\end{table}


We reduce the number of concurrent reads from BRAMs required to control one qubit. As a result, RFSoCs can drive significantly more qubits. Table~\ref{tab:gain_qubits} shows how different window sizes increase the number of qubits that can be controlled by an RFSoC based controller. This gain is independent of the ratio of FPGA clock frequency and DAC sampling rate as long as the clock frequency ratio is a multiple of the window size. If the ratio is not a multiple of the window, then the gain will be slightly lower. As an example, consider the case where the ratio is 6x. For such a system, uncompressed waveforms would require interleaving of samples between six BRAMs per waveform. If compressed waveforms with WS=8 were used, the number of BRAMs needed per waveform would be three (assuming the same distribution as shown in Figure~\ref{fig:max_samples}). The gain in the number of qubits for such a system would be 2x, which is slightly less than the case where the ratio is 8x.
The ratio between the DAC and FPGA was 16x in QICK due to which it can theoretically support about 36 qubits. Using {\sc compaqt} with WS=8, number of qubits can be increased to about 95 qubits, and for WS=16, we can drive 191 qubits concurrently. Although the availability of DACs and the limit of Frequency Division Multiplexing (FDM) may result in lesser qubits, the theoretical bounds of the control computer will no longer bottleneck the scalability of the FPGA platforms.

\subsection{Adaptive Decompression with COMPAQT}
\label{sec:adaptive}
\begin{figure}[t]
    \centering
    \includegraphics[width=\linewidth]{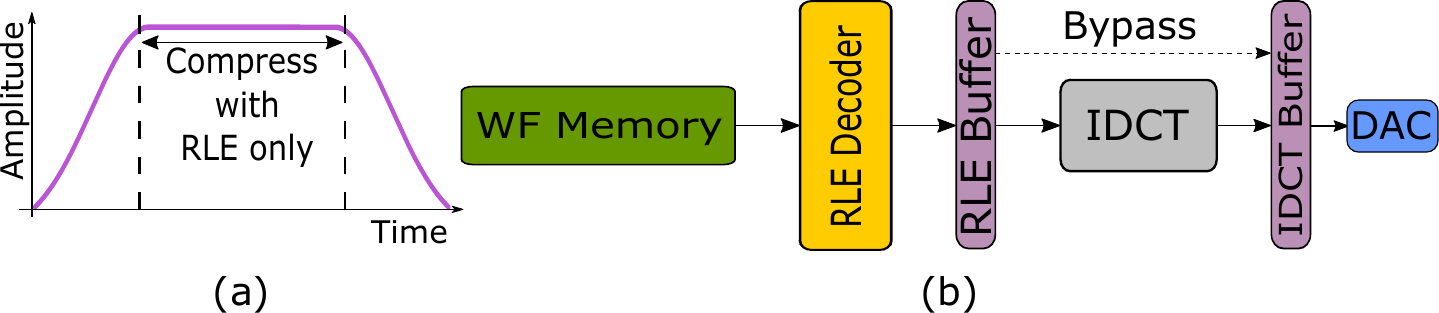}
    \caption{(a) A flat-top waveform -- the constant part of the waveform can be compressed effectively with RLE only; (b) Adaptive decompression -- the IDCT engine can be bypassed for such flat-top waveforms to save power.}
    \label{fig:adaptive}
\end{figure}

A diverse set of waveforms are used to manipulate qubits. Some gate waveforms have more structure than others. For example, the flat-top waveform as shown in Figure~\ref{fig:adaptive}(a) is commonly used to perform multi-qubit gates~\cite{Rigetti2010, Chow2011, rol2019fast, boissonneault2009dispersive}. Optimizing the decompression pipeline for such waveforms can help reduce power dissipation for ASIC designs even further. The constant period of the waveform just repeats the same value for a long duration, which can be represented by a single RLE word that can be decoded and fed directly into the buffer preceding the DAC as shown in Figure~\ref{fig:adaptive}(b). This saves power since both the memory and IDCT engine will not be used during this period. Once the flat period is over, the IDCT-driven decompression can resume.
The algorithm for adaptive decompression will treat the constant period of the waveform as a single window rather than dividing it into multiple windows (depending on the window size).



\section{Methodology}

{\noindent \textbf{Benchmarks.}}
The benchmarks used for evaluating compressed waveforms are listed in Table~\ref{tab:benchmarks}. The surface code benchmarks~\cite{Tomita2014} were selected to evaluate the effect on scalability when compressed waveforms are used, and the fidelity benchmarks were selected to complement the RB experiments summarized in Section \ref{sec:compression}. Benchmarks such as \texttt{qft-4}, \texttt{adder-4}, and \texttt{qaoa-6} include sequences of non-Clifford gates that are not used in RB experiments. Larger benchmarks using more qubits were not used since they are prone to higher variability in the results.

\begin{table}[hpt]
\caption{Benchmarks for evaluations.}
\label{tab:benchmarks}
\begin{center}
\begin{small}
\setlength{\tabcolsep}{0.05cm} 
\renewcommand{\arraystretch}{1.2}
\scalebox{0.82}{
\begin{tabular}{ | c | c | c | c | c | }
\hline
    \textbf{Benchmark}    &           \textbf{Description}                       & \textbf{\# qubits} & \textbf{Purpose} & \textbf{\# CNOTs} \\ \hline \hline
    \texttt{swap}         &           Swap gate                                  &  2                 & Fidelity         &   3 \\ \hline
    \texttt{toffoli}      &           Toffoli gate                               &  3                 & Fidelity         &  12 \\ \hline
    \texttt{qft-4}        &           Quantum Fourier Transform                  &  4                 & Fidelity         &  27 \\ \hline
    \texttt{adder-4}      &           4-bit Full Adder                           &  4                 & Fidelity         &  33 \\ \hline
    \texttt{bv-5}         &           Bernstein Vazirani                         &  6                 & Fidelity         &  2  \\ \hline
    \texttt{qaoa-6}       &           Optimization Algorithm     &  6                 & Fidelity         & 142 \\ \hline
    \texttt{qaoa-8a}      &            Optimization Algorithm     &  8                 & Fidelity         & 76 \\ \hline
    \texttt{qaoa-8b}      &            Optimization Algorithm     &  8                 & Fidelity         & 113 \\ \hline
    \texttt{qaoa-10}      &            Optimization Algorithm     &  10                & Fidelity         & 138 \\ \hline
    \texttt{surface-17}   &           Surface Code Patch (d=3)                           &  17                & Scalability      &  -   \\ \hline
    \texttt{surface-25}   &            Surface Code Patch (d=3)                           &  25                & Scalability      &  -   \\ \hline
    
\end{tabular}
}
\end{small}
\end{center}
\end{table}
    
\vspace{0.05in}
{\noindent \textbf{Software System.}}
The compression module was developed using Python and integrated with Qiskit and Qiskit Pulse~\cite{Qiskit, Alexander2020}. \texttt{DCT-N} and \texttt{DCT-W} were adapted from SciPy~\cite{scipy} while \texttt{int-DCT-W} were developed based on the HEVC transform~\cite{sze2014high}. The standard Qiskit transpiler was used to transpile the fidelity benchmarks for the target system.

\vspace{0.05in}
{\noindent \textbf{Control Hardware.}}
The hardware overhead of using DCT based compression on RFSoCs was evaluated using Xilinx's Vivado Design Suite. The QICK~\cite{qick}\footnote{\url{https://github.com/openquantumhardware/qick}} was synthesized as a baseline design on the \texttt{zc7u7ev} SoC and the \texttt{int-DCT-W} IDCT engine for different window sizes was implemented using Verilog and integrated with QICK. We validated the decompression pipeline integrated with QICK to ensure functional correctness. Furthermore, we measure the impact of integrated IDCT engine on clock frequency and resource utilization of overall control processor. To evaluate our ASIC design, we use Synopsys Design Compiler with the 40nm TSMC CLN40G cell library to estimate the power dissipation of IDCT engine, along with the Destiny cache model~\cite{poremba2015destiny} which integrates CACTI~\cite{chen2012cacti} to estimate power dissipation of SRAM based waveform memory. 

\vspace{0.05in}
{\noindent \textbf{Quantum Hardware.}}
RB and benchmark fidelity experiments were performed on real quantum hardware. RB experiments were performed on the 5-qubit IBM Bogota, 16-qubit IBM Guadalupe, and 27-qubit IBM Hanoi~\cite{ibm_systems}. Benchmark fidelities with compressed and uncompressed waveforms were evaluated on IBM Guadalupe and Toronto hardware. We evaluate fidelity (F) by computing Total Variational Distance (TVD) between ideal ($P$) and experimental ($Q$)  output distributions as shown in Equation~\ref{eq:fidelity}.

\begin{equation}
F (P,Q)= 1- TVD(P,Q)
\label{eq:fidelity}
\end{equation}

\section{Evaluations}

In this section, we evaluate 1) the impact on benchmark fidelity when compressed waveforms are used, 2) the scalability of RFSoC and ASIC controllers for important applications like Quantum Error Correction (QEC), and 3) the execution time for compressing waveforms at compile time.


\subsection{Gate Compressibility with COMPAQT}
\label{sec:gate_compressibility}

\begin{figure}
    \centering
    \includegraphics[width=1.0\linewidth]{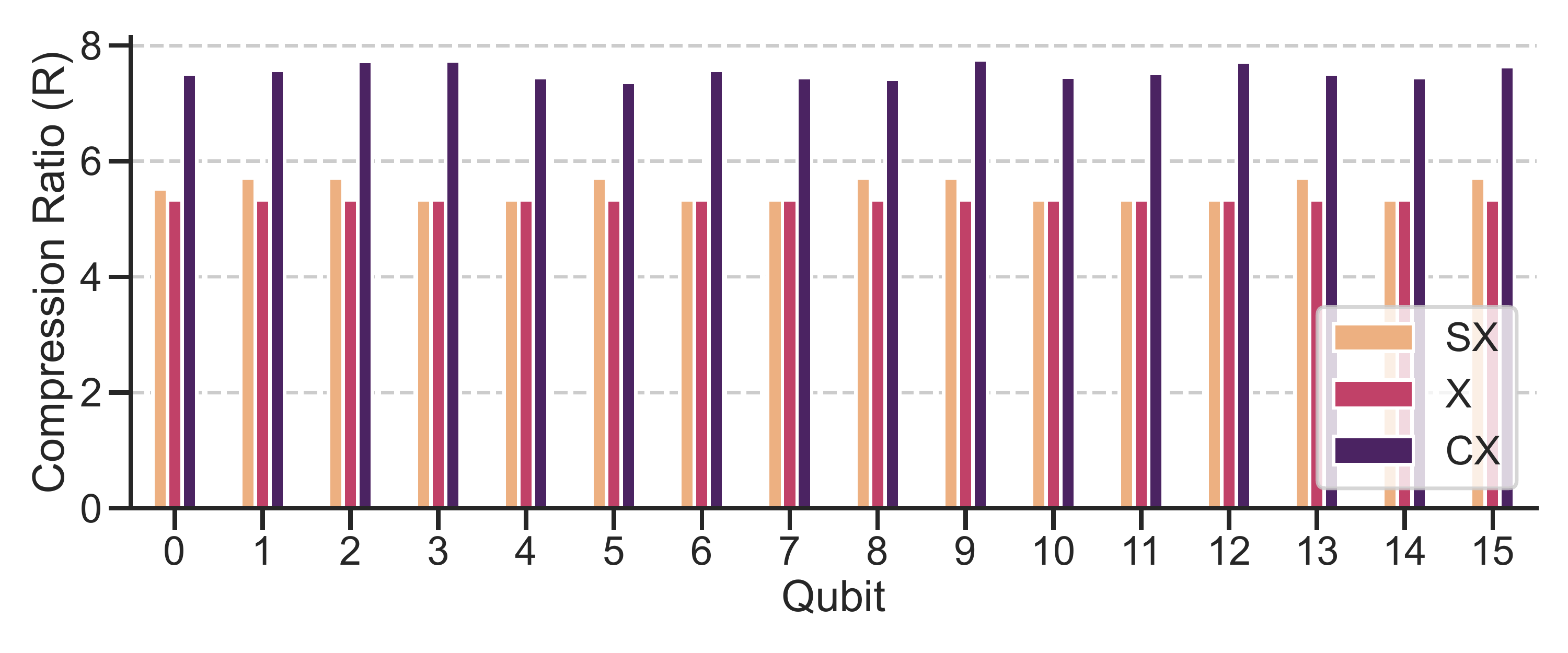}
    \caption{Compression ratios of the basis gates for all 16 qubits of IBM Guadalupe using int-DCT-W with WS=16.  }
    \label{fig:per_qubit_compression}
\end{figure}

We evaluated the compressibility of the gate pulses using {\sc compaqt} for five IBM quantum computers. Table~\ref{tab:comp} reports the minimum, average, and maximum compressibility of a control pulse used on IBM hardware. Despite the diversity in the pulse shapes, we observe that all control pulses are highly compressible. For example, as shown in Figure~\ref{fig:per_qubit_compression}, on average, we can compress the control pulses by more than 5x for each qubit device on the IBM-Guadalupe. Note that in Figure~\ref{fig:per_qubit_compression}, the compression ratios for CNOT gates are averaged over all CNOT gates that can be performed using a qubit device, whereas Table~\ref{tab:comp} reports the minimum and maximum compression ratio for an individual gate pulse on the IBM hardware. We observe that on all platforms the SX-gate has the lowest compression ratio of 5.33 with our design.  
\begin{table}[h]
\caption{Minimum, maximum, and average compression ratios with int-DCT-W (WS=16).}
\label{tab:comp}
\begin{center}
{
\setlength{\tabcolsep}{0.05cm} 
\renewcommand{\arraystretch}{1.2}
\scalebox{0.85}{
\begin{tabular}{|c|ccc|}
\hline
\multirow{2}{*}{\textbf{Machine}} & \multicolumn{3}{c|}{\textbf{Compression Ratio (R)}}                                     \\ \cline{2-4} 
                                  & \multicolumn{1}{c|}{\textbf{Min.}} & \multicolumn{1}{c|}{\textbf{Max.}} & \textbf{Avg.} \\ \hline
IBM Toronto                       & \multicolumn{1}{c|}{5.33}          & \multicolumn{1}{c|}{8.11}          & 6.49          \\ \hline
IBM Montreal                      & \multicolumn{1}{c|}{5.33}          & \multicolumn{1}{c|}{8.02}          & 6.45          \\ \hline
IBM Mumbai                        & \multicolumn{1}{c|}{5.33}          & \multicolumn{1}{c|}{8.05}          & 6.47          \\ \hline
IBM Guadalupe                     & \multicolumn{1}{c|}{5.33}          & \multicolumn{1}{c|}{8.02}          & 6.48          \\ \hline
IBM Lima                          & \multicolumn{1}{c|}{5.33}          & \multicolumn{1}{c|}{7.92}          & 6.33          \\ \hline
\end{tabular}
}}

\end{center}
\end{table}

\subsection{Gate Fidelity with COMPAQT}
To evaluate whether compressed waveforms degrade the fidelity of quantum circuits, we run circuits for 80K shots with and without waveform compression and compute the normalized fidelity by taking the ratio of the fidelities of {\sc compaqt} and the uncompressed baseline.  

\begin{figure}[t]
    \centering
    \includegraphics[width=1.00\linewidth]{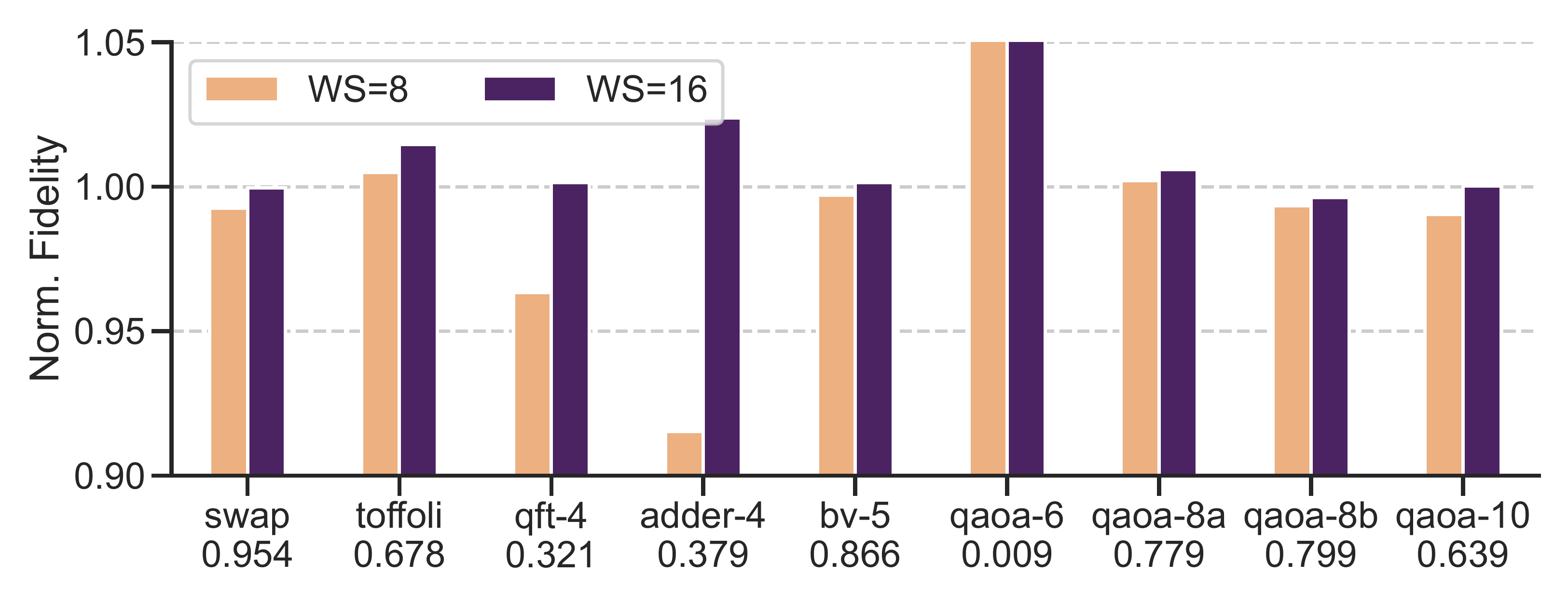}
    \caption{{\sc compaqt} fidelity normalized to baseline fidelity on IBM Guadalupe. The baseline fidelities are annotated beneath the benchmark names.}
    \label{fig:bm_fidelities}
\end{figure}

Figure \ref{fig:bm_fidelities} shows the normalized benchmark fidelities for WS=8 and WS=16 with \texttt{int-DCT-W} on IBM Guadalupe. Norm. fidelity close to one means both the baseline and {\sc compaqt} have identical fidelities, indicating no degradation due to lossy compression. For window size of WS=16 we see no fidelity degradation but for the WS=8, there are significant fidelity losses for some benchmarks due to distortions introduced at the boundaries of consecutive windows. These distortions can be reduced by using overlapping windows to compress the waveform. Note that in some cases normalized fidelity is slightly greater than one, indicating {\sc compaqt} is more reliable. This can be attributed to the variability in the experiments. Note that the fidelity for the QAOA benchmarks was determined by computing the normalized fidelity~\cite{hashim2021randomized, lubinski2021application}. We observe that the fidelity degradation is less than 0.5\% for the largest QAOA benchmark using 10 qubits with a window size of 16.


\subsection{Scalability of COMPAQT with RFSoC}

To evaluate the scalability of {\sc compaqt} on RFSoC platforms, we evaluate the maximum clock frequency, resource utilization and maximum number of concurrent gates that can be executed for key benchmark circuits.

\begin{figure}[b]
    \centering
    \includegraphics[width=1\linewidth]{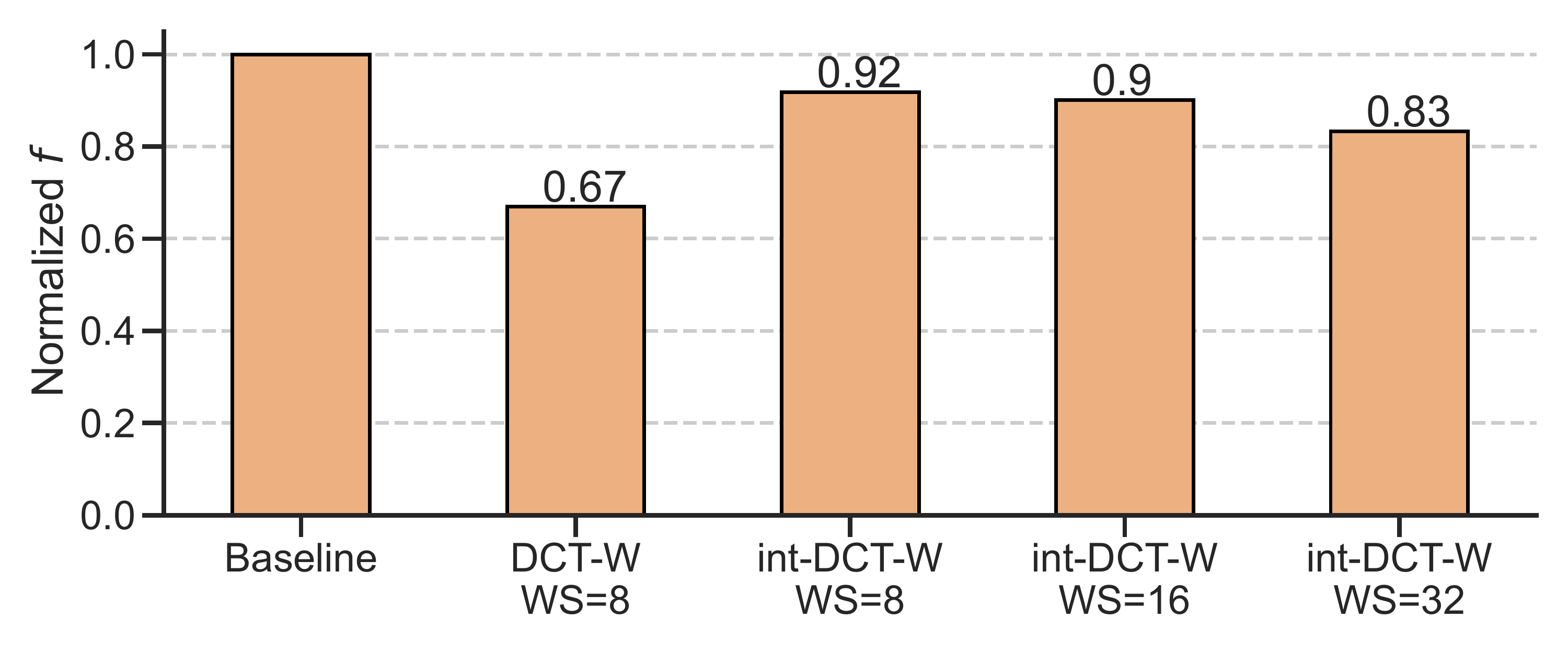}
    \caption{Degradation of clock frequency compared to baseline for different decompressed waveform memory architectures.}
    \label{fig:freq}
\end{figure}

{\noindent \textbf{Clock Frequency.} }We use QICK~\cite{qick} as the baseline design with a maximum frequency of 294 MHz.  For {\sc compaqt}, the maximum achievable frequency drops by more than 33\% to 195 MHz for the pipelined \texttt{DCT-W} engine, as shown in Figure \ref{fig:freq}. This degradation results from complex design and long critical path due to multipliers. In comparison, the unpipelined \texttt{int-DCT-W} design introduced a worst-case degradation of 10\%, which can be pipelined to enable a design with no clock frequency degradation. These results can be scaled to arbitrary number of qubits as to control multiple qubits the individual qubit control block will be instantiated in parallel having a little to no effect on the critical path.

\begin{table}[hpt]
\caption{FPGA resource usage for the baseline and a single IDCT engine for int-DCT-W for different window sizes. The percentages represent the percent utilization of the total resources. The designs were synthesized for the same SoC - Xilinx zc7u7ev.}
\label{tab:resources}
\begin{center}
\begin{small}
\setlength{\tabcolsep}{0.05cm} 
\renewcommand{\arraystretch}{1.2}
\scalebox{0.9}{
\begin{tabular}{|c|c|c|}
\hline
\textbf{Design}           & \textbf{LUT Utilization}         & \textbf{FF Utilization}    \\ \hline \hline 
Baseline                   & 3386 (1.4\%)          & 6448 (1.4\%)    \\ \hline 
\texttt{int-DCT-W} (WS=8)  & 601 (0.26\%)          & 266 (0.05\%)    \\ \hline 
\texttt{int-DCT-W} (WS=16) & 1954 (0.85\%)         & 671 (0.15\%)    \\ \hline 
\texttt{int-DCT-W} (WS=32) & 9063 (3.93\%)         & 1197 (0.26\%) \\ \hline
\end{tabular}
}

\end{small}
\end{center}
\end{table}


\vspace{0.05in}
{\noindent \textbf{Resource Utilization.}} Table \ref{tab:resources} shows the usage of critical FPGA resources like Look-Up Tables (LUTs) and Flip-Flops (FFs) of the baseline design (QICK) and a single IDCT engine for \texttt{int-DCT-W} with different window sizes. The baseline design can control a single qubit, but it includes components like the AXI interface that will not scale linearly with the number of qubits. The resources are not enough to accommodate 100+ qubits. The overall resource usage per module will be lower than 1\% for RFSoCs, as they have significantly more resources than the device used for this evaluation (the RFSoC used by QICK has 400K+ LUTs and 800K+ FFs). Compared to BRAMs, LUTs and FFs are more readily available on FPGA and so using a compressed memory architecture trades memory units (BRAMs) with system logic (compute) per qubit in FPGAs. The design of the IDCT hardware used for these evaluations can be further optimized by pipelining. 

Although increasing window size can improve compression efficiency, using a larger window size such as WS=32 with \texttt{int-DCT-W} uses significantly more resources than WS=16, which makes it a sub-optimal design as a single iDCT engine uses more than 4\% of the total LUTs. In summary, int-DCT with WS=32 reduces clock frequency and adds complexity, limiting the scalability.


\vspace{0.05in}
{\noindent \textbf{Scalability of QEC Experiments.}} Demonstrating a small set of logical qubits that run quantum error correction is a crucial milestone in achieving fault-tolerant quantum computation. Surface code is a leading candidate for implementing QEC~\cite{Fowler2012} and its effectiveness have been demonstrated experimentally for low-distance codes~\cite{krinner2021realizing, marques2022logical, zhao2021realizing, chen2021}. As systems scale and add more qubits to move towards fault-tolerant systems, the number of logical qubits that a single controller can support will thus be an important metric. 

\begin{figure}[t]
    \centering 
    \includegraphics[width=\linewidth]{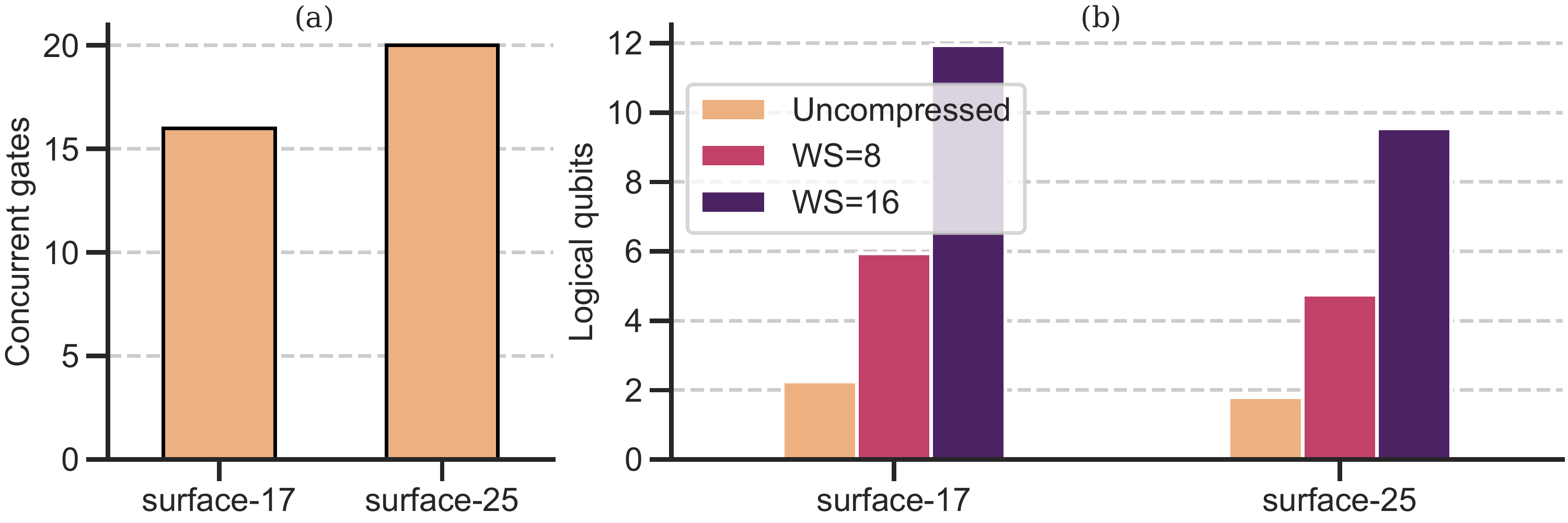}
    \caption{Increasing the number of logical qubits that can be supported; (a) The peak concurrent operations for a d=3 code; (b) The maximum logical qubits that can be supported by different Xilinx RFSoC based designs.}
    \label{fig:surface}
\end{figure}

Figure~\ref{fig:surface}(a) shows the maximum number of concurrent operations at any given time during the syndrome generation cycle in two variants of a distance three surface code patch. More than 80\% of the physical qubits to create one logical qubit are driven concurrently, which makes it important for a controller hardware to be able to support concurrency in as many qubits as possible. Figure~\ref{fig:surface}(b) show that {\sc compaqt} can control 5x more logical qubits compared to an uncompressed baseline. Similarly, for ASIC controllers, using compressed waveform memory can reduce memory power dissipation by more than 2x which can allow more logical qubits to be controlled. 

\subsection{Scalability of COMPAQT with ASICs}

\begin{figure}[b]
    \centering
    \includegraphics[width=\linewidth]{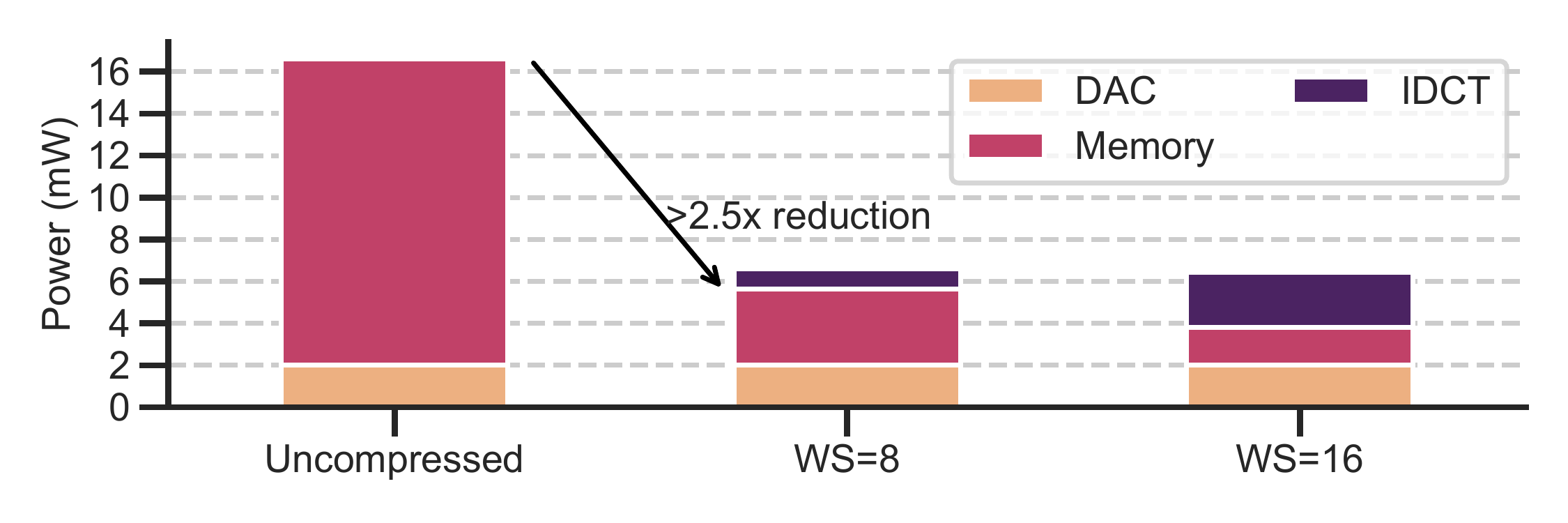}
    \caption{Power dissipated by the cryogenic controller with the uncompressed and compressed memory.}
    \label{fig:sram_power}
\end{figure}

The intrinsic BRAM bandwidth on FPGA-fabric limits their scalability. Compressing waveform memory can help improve such systems. However, unlike BRAMs, SRAMs used in ASICs can operate at much higher frequencies and deliver significantly more bandwidth. However, for ASICs designed to work at cryogenic temperatures, it is crucial for the overall architecture to be as power efficient as possible due to tiny power budget at 4K. While the cooling capacity of dilution refrigerators will probably have to be expanded in the future to accommodate more qubits, the power budget will still be limited and minimizing the power dissipated by digital components like waveform memory can help optimize other components of qubit control that are critical for high-fidelity gates. For example, 50\% of the total power (23mW) in the cryogenic chip presented in~\cite{Frank2022} was dissipated by digital components (including the digital frontend of the DACs).

In general, the energy dissipated per access by memory increases with the size of the memory. Since SRAMs can operate at much higher frequencies than FPGA BRAMs, interleaving memory samples between multiple banks is not necessary and so every window can have variable number of compressed samples since every sample will be fetched sequentially and decompression will start every time an RLE codeword is detected. By compressing waveforms, SRAMs can be made smaller and fewer accesses are required to read the same waveform when compared to an uncompressed waveform memory. Figure \ref{fig:sram_power} shows how memory power is reduced by more than 2.5x when compressed waveforms are used for a CMOS based cryogenic controller. Memory power was estimated with the Destiny model~\cite{poremba2015destiny}. The DAC power of 2mW was added as a reference and the power dissipated by the \texttt{int-DCT-W} IDCT engine for corresponding window sizes was estimated via Synopsys Design Compiler with a TSMC 40nm cell library -- actual power dissipated by the IDCT engine can be even lower, the data shown here motivates the use of compression as the overhead of using the IDCT engine does not overshadow the decrease in memory power. Using the compressed waveform architecture can thus reduce the overall power dissipated by the waveform memory by at least 3x.

\begin{figure}[t]
    \centering
    \includegraphics[width=0.9\linewidth]{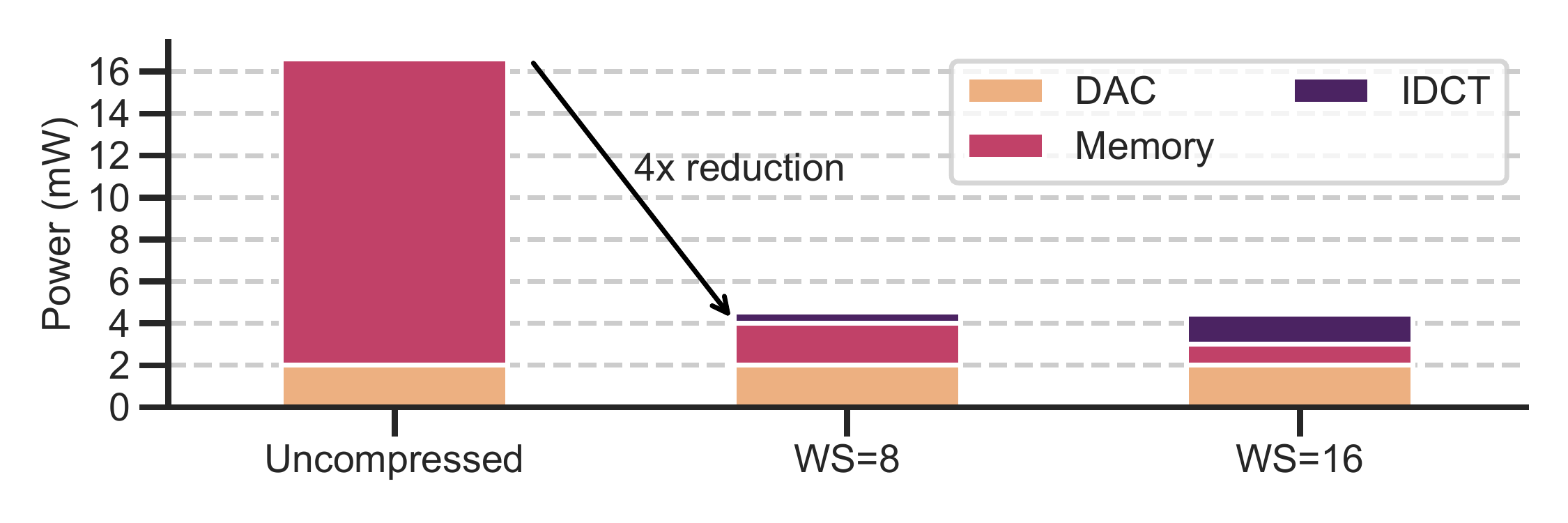}
    \caption{Power consumption for flat-top waveform with uncompressed baseline and COMPAQT using adaptive decompression with WS=8 and WS=16.}
    \label{fig:adaptive_power}
\end{figure}

Moreover, we can further reduce power consumption using adaptive decompression as shown in Figure \ref{fig:adaptive_power}. Since memory is accessed only during the rise and fall ramps of the waveform, memory power is reduced over the entire period along with the power dissipated by the IDCT engine, yielding an additional 1.5x power savings. This gain will vary with the duration of the flat-top waveform, a 100ns waveform was used for this evaluation.

\subsection{Software Overhead of COMPAQT}
\begin{figure}[t]
    \centering
    \includegraphics[width=\linewidth]{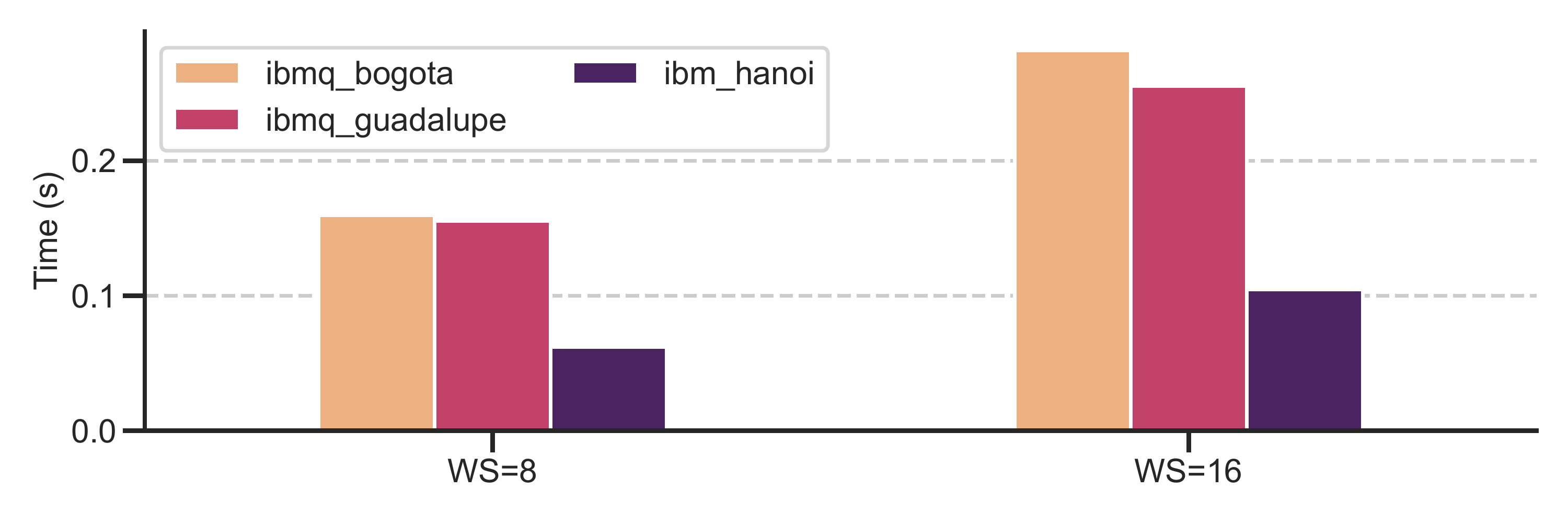}
    \caption{Average time to compress a gate waveform using int-DCT-W compression.}
    \label{fig:exec_time}
\end{figure}

To evaluate the overhead of decompression, we measure the compression latency of the {\sc compaqt} compiler module.
Figure \ref{fig:exec_time} shows the average time to compress a waveform using \texttt{int-DCT-W} for three IBM machines. 
Since waveforms need only be compressed at the end of every calibration cycle, the overhead of compressing waveforms is negligible compared to the time taken to calibrate all qubits -- regular calibration of the Sycamore chip lasts four hours~\cite{Arute2019}. As a result, compression has little or no overhead on the system as a whole and does not affect system performance metrics such as repetition rate and Circuit Level Operations Per Second (CLOPS)~\cite{clops}.
 

\section{Related Work}


Efficient qubit control is crucial to scale quantum computers. While most control platforms use discrete FPGAs and analog ICs, we expect RFSoC-based and custom ASIC-based solutions to grow in the future.

\vspace{0.05in}
{\noindent \textbf{RFSoC controllers.}}
We use QICK as the baseline, which is an open source RFSoC-based qubit control framework~\cite{qick}, that reduces discrete analog components by using direct digital synthesis. Similarly, recent works from a different group of researchers also demonstrate the effectiveness of RFSoCs in controlling superconducting qubits~\cite{icarus,gebauerflexible}.

\vspace{0.05in}
{\noindent \textbf{ASIC controllers.}}
Studies on ASIC controllers for qubit control have mainly been focused on developing cryogenic controllers. Fully/semi-autonomous cryogenic chips were shown by~\cite{Frank2022, Kang2022, VanDijk2020_} that integrated both digital and analog blocks on a single ASIC. These prototype chips are designed to control a small number of qubits. Other CMOS based ASICs were presented in~\cite{9365762, kang2021, Xue2021, Pauka2021, Bardin2019, vanDijk2020, Kiene2021}. A complete system-level study for a control processor using SFQ pulses was presented in~\cite{digiQ}. Other controller architectures that use SFQ pulses were described in~\cite{mcdermott2014, McDermott2018, liebermann2016, Li2019, leonard2019}. 

\vspace{0.05in}
{\noindent \textbf{Control architectures.}}
Many prior works are focused on FPGA/RFSoC/ASIC controllers. Fu et al. presented an FPGA-based microarchitecture for controlling a transmon qubit~\cite{Fu2017, Fu2016}. Scalable architectures for QECC applications were presented in~\cite{das2020scalable, Tannu2017}, whereas ISA specific works for qubit control include~\cite{Fu2019} which introduced an executable Quantum ISA, \cite{Butko2020} which compared their proposed scalar and vector instruction sets with other ISAs, and \cite{Lao2021} which recommended using a small number of expressive two-qubit gates. At the same time, several works focused on microarchitectural challenges in building a quantum computer with hundreds of thousands of qubits ~\cite{Meter2006, Metodi2011, VanMeter2013}. 

\section{Discussion}
\label{sec:discussion}

Qubit control pulses have a tight footprint in the frequency domain. Any spurious frequencies in the control pulse can introduce control error, crosstalk, and leakage errors. As a result, due to the tiny footprint in the frequency domain, control pulses can be compressed and stored efficiently. To the best of our knowledge, {\sc compaqt} is the first to leverage this fundamental property to build a scalable and efficient control using compressed waveform memories and compiler-guided compression that ensures high fidelity. Moreover, these insights can be used for designing single flux quantum (SFQ) based qubit control, in which on-chip memory is limited to tens of kilobytes~\cite{digiQ}. 

{\noindent \textbf{Enabling Novel Design Tradeoffs.}} We expose new tradeoffs to enable scalable qubit control. A key novelty of the proposed microarchitecture is the hardware software co-design of IDCT engine that offers high compressibility without reducing the gate fidelity. To achieve this, we designed a discrete cosine transform-based compression scheme that uses only integer arithmetic and eliminates the use of complex multiplier circuits. However, naive approximations in the DCT engine can result in distortions in certain gate pulses. To avoid such corner-cases, we propose compiler-guided compression to navigate the trade-off between compressibility and fidelity. Our design enables the lightweight design of the IDCT engine to ensure efficient synthesis on current RFSoC-based control computers. Moreover, we empirically show that {\sc compaqt} can support 5x more qubits compared to the uncompressed baseline with minimal degradation in gate and circuit fidelity.

{\noindent \textbf{Impact on Future Designs.}} To understand the applicability of our insights on future qubit architectures, we evaluate the compressibility of control pulses for current and emerging qubit technologies. Table~\ref{tab:diss} shows that control pulses are compressible not only for mainstream transmon qubits using single and two qubit gates but also complex control pulses along with the control pulses for emerging fluxonium qubits. To enable fault-tolerant quantum computers, we must focus on microarchitectural challenges in building control hardware that can support thousands of qubits. To that end, we propose {\sc compaqt}, which can help mitigate the bottlenecks in the qubit control pipeline by leveraging compressibility of control pulses.

\begin{table}[hpt]
\caption{Compression Ratios (R) for complex gate pulses for transmon and single qubit gates for emerging fluxonium using int-DCT-W with WS=16.}
\label{tab:diss}

\begin{center}
\setlength{\tabcolsep}{0.075cm} 
\renewcommand{\arraystretch}{1.3}
\scalebox{0.95}{
\begin{tabular}{ | c | c | c | c | }
\hline
    \textbf{Device Type}    &           \textbf{Gate}                & \textbf{Description} & \textbf{R}  \\ \hline \hline
    Transmon      &    \texttt{iToffoli} &     Three Qubit Gate Pulse~\cite{kim2022high}    &  8.32  \\ \hline 
    Transmon      &    \texttt{Toffoli} &     Three Qubit Gate Pulse~\cite{zahedinejad2016designing}    &  5.31 \\ \hline
    Transmon  &    \texttt{CCZ}      & Three Qubit Gate Pulse~\cite{zahedinejad2016designing}     & 5.59  \\ \hline
    Fluxonium  &    \texttt{X},$\frac{X}{2}$,$\frac{Z}{2}$,$\frac{Y}{2}$          & Single Qubit Gate Pulse~\cite{propson2022robust}  & 7.2  \\ \hline
    
\end{tabular}
}
\end{center}
\end{table}



\section{Conclusions}

As qubits increase beyond the 100-qubit mark, scaling the control hardware with brute-force methods will be infeasible. Integration using RFSoCs and cryogenic ASICs will be required to enable scale qubit control hardware. However, in this paper, we show that the internal memory bandwidth used for playing waveforms in RFSoC-based systems becomes a bottleneck. Furthermore, for controller ASICs, the power dissipated by the waveform memory is a significant portion of the total power and optimizing the memory architecture will enable better scalability in such systems. To solve these problems with waveform memory in different types of controllers, we present {\sc compaqt}, a compressed waveform memory architecture that compresses waveforms in software before they are loaded on the controller. In hardware, waveforms are decompressed just before they are needed. Our results show that for RFSoC-based systems, the number of qubits that can be supported increases by more than 5x compared to the baseline. For CMOS ASICs, the reduced number of accesses required per waveform reduces power dissipation by more than 2.5x. Compressed waveform memory has minimal effect on the gate fidelity, as shown on real quantum hardware, while enabling more scalable controllers.
\section*{Acknowledgments}
The authors thank Chaithanya Naik Mude for the technical discussions and anonymous reviewers for their feedback. This research was partially supported by the Vice Chancellor Office for Research and Graduate Education at the University of Wisconsin–Madison with funding from the Wisconsin Alumni Research Foundation.




%



\bibliographystyle{IEEEtranS}
\bibliography{IEEEabrv, main}

\begin{thebibliography}{10}
\providecommand{\url}[1]{#1}
\csname url@samestyle\endcsname
\providecommand{\newblock}{\relax}
\providecommand{\bibinfo}[2]{#2}
\providecommand{\BIBentrySTDinterwordspacing}{\spaceskip=0pt\relax}
\providecommand{\BIBentryALTinterwordstretchfactor}{4}
\providecommand{\BIBentryALTinterwordspacing}{\spaceskip=\fontdimen2\font plus
\BIBentryALTinterwordstretchfactor\fontdimen3\font minus
  \fontdimen4\font\relax}
\providecommand{\BIBforeignlanguage}[2]{{%
\expandafter\ifx\csname l@#1\endcsname\relax
\typeout{** WARNING: IEEEtranS.bst: No hyphenation pattern has been}%
\typeout{** loaded for the language `#1'. Using the pattern for}%
\typeout{** the default language instead.}%
\else
\language=\csname l@#1\endcsname
\fi
#2}}
\providecommand{\BIBdecl}{\relax}
\BIBdecl

\bibitem{ahmed1974discrete}
N.~Ahmed, T.~Natarajan, and K.~R. Rao, ``Discrete cosine transform,''
  \emph{IEEE transactions on Computers}, vol. 100, no.~1, pp. 90--93, 1974.

\bibitem{alameldeen2004adaptive}
A.~R. Alameldeen and D.~A. Wood, ``Adaptive cache compression for
  high-performance processors,'' in \emph{Proceedings. 31st Annual
  International Symposium on Computer Architecture, 2004.}\hskip 1em plus 0.5em
  minus 0.4em\relax IEEE, 2004, pp. 212--223.

\bibitem{Alexander2020}
\BIBentryALTinterwordspacing
T.~Alexander, N.~Kanazawa, D.~J. Egger, L.~Capelluto, C.~J. Wood,
  A.~Javadi-Abhari, and D.~C. McKay, ``Qiskit pulse: programming quantum
  computers through the cloud with pulses,'' \emph{Quantum Science and
  Technology}, vol.~5, no.~4, p. 044006, Aug. 2020. [Online]. Available:
  \url{https://doi.org/10.1088/2058-9565/aba404}
\BIBentrySTDinterwordspacing

\bibitem{chen2021}
\BIBentryALTinterwordspacing
and Zijun~Chen, K.~J. Satzinger, J.~Atalaya, A.~N. Korotkov, A.~Dunsworth,
  D.~Sank, C.~Quintana, M.~McEwen, R.~Barends, P.~V. Klimov, S.~Hong, C.~Jones,
  A.~Petukhov, D.~Kafri, S.~Demura, B.~Burkett, C.~Gidney, A.~G. Fowler,
  A.~Paler, H.~Putterman, I.~Aleiner, F.~Arute, K.~Arya, R.~Babbush, J.~C.
  Bardin, A.~Bengtsson, A.~Bourassa, M.~Broughton, B.~B. Buckley, D.~A. Buell,
  N.~Bushnell, B.~Chiaro, R.~Collins, W.~Courtney, A.~R. Derk, D.~Eppens,
  C.~Erickson, E.~Farhi, B.~Foxen, M.~Giustina, A.~Greene, J.~A. Gross, M.~P.
  Harrigan, S.~D. Harrington, J.~Hilton, A.~Ho, T.~Huang, W.~J. Huggins, L.~B.
  Ioffe, S.~V. Isakov, E.~Jeffrey, Z.~Jiang, K.~Kechedzhi, S.~Kim, A.~Kitaev,
  F.~Kostritsa, D.~Landhuis, P.~Laptev, E.~Lucero, O.~Martin, J.~R. McClean,
  T.~McCourt, X.~Mi, K.~C. Miao, M.~Mohseni, S.~Montazeri, W.~Mruczkiewicz,
  J.~Mutus, O.~Naaman, M.~Neeley, C.~Neill, M.~Newman, M.~Y. Niu, T.~E.
  O'Brien, A.~Opremcak, E.~Ostby, B.~Pat{\'{o}}, N.~Redd, P.~Roushan, N.~C.
  Rubin, V.~Shvarts, D.~Strain, M.~Szalay, M.~D. Trevithick, B.~Villalonga,
  T.~White, Z.~J. Yao, P.~Yeh, J.~Yoo, A.~Zalcman, H.~Neven, S.~Boixo,
  V.~Smelyanskiy, Y.~Chen, A.~Megrant, and J.~Kelly, ``Exponential suppression
  of bit or phase errors with cyclic error correction,'' \emph{Nature}, vol.
  595, no. 7867, pp. 383--387, Jul. 2021. [Online]. Available:
  \url{https://doi.org/10.1038/s41586-021-03588-y}
\BIBentrySTDinterwordspacing

\bibitem{Qiskit}
M.~S. ANIS, H.~Abraham, AduOffei, R.~Agarwal, G.~Agliardi, M.~Aharoni, I.~Y.
  Akhalwaya, G.~Aleksandrowicz, T.~Alexander, M.~Amy, S.~Anagolum, E.~Arbel,
  A.~Asfaw, A.~Athalye, A.~Avkhadiev, C.~Azaustre, A.~Banerjee, S.~Banerjee,
  W.~Bang, A.~Bansal, P.~Barkoutsos, A.~Barnawal, G.~Barron, G.~S. Barron,
  L.~Bello, Y.~Ben-Haim, D.~Bevenius, D.~Bhatnagar, A.~Bhobe, P.~Bianchini,
  L.~S. Bishop, C.~Blank, S.~Bolos, S.~Bopardikar, S.~Bosch, S.~Brandhofer,
  Brandon, S.~Bravyi, N.~Bronn, Bryce-Fuller, D.~Bucher, A.~Burov, F.~Cabrera,
  P.~Calpin, L.~Capelluto, J.~Carballo, G.~Carrascal, A.~Carriker, I.~Carvalho,
  A.~Chen, C.-F. Chen, E.~Chen, J.~C. Chen, R.~Chen, F.~Chevallier,
  R.~Cholarajan, J.~M. Chow, S.~Churchill, C.~Claus, C.~Clauss, C.~Clothier,
  R.~Cocking, R.~Cocuzzo, J.~Connor, F.~Correa, A.~J. Cross, A.~W. Cross,
  S.~Cross, J.~Cruz-Benito, C.~Culver, A.~D. C{\'o}rcoles-Gonzales, N.~D,
  S.~Dague, T.~E. Dandachi, A.~N. Dangwal, J.~Daniel, M.~Daniels, M.~Dartiailh,
  A.~R. Davila, F.~Debouni, A.~Dekusar, A.~Deshmukh, M.~Deshpande, D.~Ding,
  J.~Doi, E.~M. Dow, E.~Drechsler, E.~Dumitrescu, K.~Dumon, I.~Duran,
  K.~EL-Safty, E.~Eastman, G.~Eberle, A.~Ebrahimi, P.~Eendebak, D.~Egger,
  A.~Espiricueta, M.~Everitt, D.~Facoetti, Farida, P.~M. Fern{\'a}ndez,
  S.~Ferracin, D.~Ferrari, A.~H. Ferrera, R.~Fouilland, A.~Frisch, A.~Fuhrer,
  B.~Fuller, M.~GEORGE, J.~Gacon, B.~G. Gago, C.~Gambella, J.~M. Gambetta,
  A.~Gammanpila, L.~Garcia, T.~Garg, S.~Garion, T.~Gates, L.~Gil, A.~Gilliam,
  A.~Giridharan, J.~Gomez-Mosquera, Gonzalo, S.~de~la Puente~Gonz{\'a}lez,
  J.~Gorzinski, I.~Gould, D.~Greenberg, D.~Grinko, W.~Guan, J.~A. Gunnels,
  N.~Gupta, J.~M. G{\"u}nther, M.~Haglund, I.~Haide, I.~Hamamura, O.~C. Hamido,
  F.~Harkins, A.~Hasan, V.~Havlicek, J.~Hellmers, {\L}.~Herok, S.~Hillmich,
  H.~Horii, C.~Howington, S.~Hu, W.~Hu, J.~Huang, R.~Huisman, H.~Imai,
  T.~Imamichi, K.~Ishizaki, Ishwor, R.~Iten, T.~Itoko, A.~Javadi,
  A.~Javadi-Abhari, W.~Javed, M.~Jivrajani, K.~Johns, S.~Johnstun,
  Jonathan-Shoemaker, JosDenmark, JoshDumo, J.~Judge, T.~Kachmann, A.~Kale,
  N.~Kanazawa, J.~Kane, Kang-Bae, A.~Kapila, A.~Karazeev, P.~Kassebaum,
  J.~Kelso, S.~Kelso, V.~Khanderao, S.~King, Y.~Kobayashi, A.~Kovyrshin,
  R.~Krishnakumar, V.~Krishnan, K.~Krsulich, P.~Kumkar, G.~Kus, R.~LaRose,
  E.~Lacal, R.~Lambert, J.~Lapeyre, J.~Latone, S.~Lawrence, C.~Lee, G.~Li,
  J.~Lishman, D.~Liu, P.~Liu, Y.~Maeng, S.~Maheshkar, K.~Majmudar, A.~Malyshev,
  M.~E. Mandouh, J.~Manela, Manjula, J.~Marecek, M.~Marques, K.~Marwaha,
  D.~Maslov, P.~Maszota, D.~Mathews, A.~Matsuo, F.~Mazhandu, D.~McClure,
  M.~McElaney, C.~McGarry, D.~McKay, D.~McPherson, S.~Meesala, D.~Meirom,
  C.~Mendell, T.~Metcalfe, M.~Mevissen, A.~Meyer, A.~Mezzacapo, R.~Midha,
  Z.~Minev, A.~Mitchell, N.~Moll, A.~Montanez, G.~Monteiro, M.~D. Mooring,
  R.~Morales, N.~Moran, D.~Morcuende, S.~Mostafa, M.~Motta, R.~Moyard,
  P.~Murali, J.~M{\"u}ggenburg, D.~Nadlinger, K.~Nakanishi, G.~Nannicini,
  P.~Nation, E.~Navarro, Y.~Naveh, S.~W. Neagle, P.~Neuweiler, A.~Ngoueya,
  J.~Nicander, Nick-Singstock, P.~Niroula, H.~Norlen, NuoWenLei, L.~J.
  O'Riordan, O.~Ogunbayo, P.~Ollitrault, T.~Onodera, R.~Otaolea, S.~Oud,
  D.~Padilha, H.~Paik, S.~Pal, Y.~Pang, A.~Panigrahi, V.~R. Pascuzzi,
  S.~Perriello, E.~Peterson, A.~Phan, F.~Piro, M.~Pistoia, C.~Piveteau,
  J.~Plewa, P.~Pocreau, A.~Pozas-Kerstjens, R.~Pracht, M.~Prokop, V.~Prutyanov,
  S.~Puri, D.~Puzzuoli, J.~P{\'e}rez, Quintiii, R.~I. Rahman, A.~Raja,
  R.~Rajeev, N.~Ramagiri, A.~Rao, R.~Raymond, O.~Reardon-Smith, R.~M.-C.
  Redondo, M.~Reuter, J.~Rice, M.~Riedemann, D.~Risinger, M.~L. Rocca, D.~M.
  Rodr{\'\i}guez, RohithKarur, B.~Rosand, M.~Rossmannek, M.~Ryu, T.~SAPV,
  A.~Saha, A.~Ash-Saki, M.~Sandberg, H.~Sandesara, R.~Sapra, H.~Sargsyan,
  A.~Sarkar, N.~Sathaye, B.~Schmitt, C.~Schnabel, Z.~Schoenfeld, T.~L.
  Scholten, E.~Schoute, M.~Schulterbrandt, J.~Schwarm, J.~Seaward, Sergi, I.~F.
  Sertage, K.~Setia, F.~Shah, N.~Shammah, R.~Sharma, Y.~Shi, J.~Shoemaker,
  A.~Silva, A.~Simonetto, D.~Singh, P.~Singh, P.~Singkanipa, Y.~Siraichi, Siri,
  J.~Sistos, I.~Sitdikov, S.~Sivarajah, M.~B. Sletfjerding, J.~A. Smolin,
  M.~Soeken, I.~O. Sokolov, I.~Sokolov, SooluThomas, Starfish, D.~Steenken,
  M.~Stypulkoski, A.~Suau, S.~Sun, K.~J. Sung, M.~Suwama, O.~S{\l}owik,
  H.~Takahashi, T.~Takawale, I.~Tavernelli, C.~Taylor, P.~Taylour, S.~Thomas,
  M.~Tillet, M.~Tod, M.~Tomasik, E.~de~la Torre, J.~L.~S. Toural, K.~Trabing,
  M.~Treinish, D.~Trenev, TrishaPe, F.~Truger, G.~Tsilimigkounakis, D.~Tulsi,
  W.~Turner, Y.~Vaknin, C.~R. Valcarce, F.~Varchon, A.~Vartak, A.~C. Vazquez,
  P.~Vijaywargiya, V.~Villar, B.~Vishnu, D.~Vogt-Lee, C.~Vuillot, J.~Weaver,
  J.~Weidenfeller, R.~Wieczorek, J.~A. Wildstrom, J.~Wilson, E.~Winston,
  WinterSoldier, J.~J. Woehr, S.~Woerner, R.~Woo, C.~J. Wood, R.~Wood, S.~Wood,
  J.~Wootton, M.~Wright, B.~Yang, D.~Yeralin, R.~Yonekura, D.~Yonge-Mallo,
  R.~Young, J.~Yu, L.~Yu, C.~Zachow, L.~Zdanski, H.~Zhang, C.~Zoufal, aeddins
  ibm, alexzhang13, b63, bartek bartlomiej, bcamorrison, brandhsn, catornow,
  charmerDark, deeplokhande, dekel.meirom, dime10, ehchen, fanizzamarco,
  fs1132429, gadial, galeinston, georgezhou20, georgios ts, gruu, hhorii,
  hykavitha, itoko, jliu45, jscott2, klinvill, krutik2966, ma5x, michelle4654,
  msuwama, ntgiwsvp, ordmoj, sagar pahwa, pritamsinha2304, ryancocuzzo, saswati
  qiskit, septembrr, sethmerkel, shaashwat, sternparky, strickroman, tigerjack,
  tsura crisaldo, welien, willhbang, yang.luh, and M.~{\v{C}}epulkovskis,
  ``Qiskit: An open-source framework for quantum computing,'' 2021.

\bibitem{Arute2019}
\BIBentryALTinterwordspacing
F.~Arute, K.~Arya, R.~Babbush, D.~Bacon, J.~C. Bardin, R.~Barends, R.~Biswas,
  S.~Boixo, F.~G. S.~L. Brandao, D.~A. Buell, B.~Burkett, Y.~Chen, Z.~Chen,
  B.~Chiaro, R.~Collins, W.~Courtney, A.~Dunsworth, E.~Farhi, B.~Foxen,
  A.~Fowler, C.~Gidney, M.~Giustina, R.~Graff, K.~Guerin, S.~Habegger, M.~P.
  Harrigan, M.~J. Hartmann, A.~Ho, M.~Hoffmann, T.~Huang, T.~S. Humble, S.~V.
  Isakov, E.~Jeffrey, Z.~Jiang, D.~Kafri, K.~Kechedzhi, J.~Kelly, P.~V. Klimov,
  S.~Knysh, A.~Korotkov, F.~Kostritsa, D.~Landhuis, M.~Lindmark, E.~Lucero,
  D.~Lyakh, S.~Mandr{\`{a}}, J.~R. McClean, M.~McEwen, A.~Megrant, X.~Mi,
  K.~Michielsen, M.~Mohseni, J.~Mutus, O.~Naaman, M.~Neeley, C.~Neill, M.~Y.
  Niu, E.~Ostby, A.~Petukhov, J.~C. Platt, C.~Quintana, E.~G. Rieffel,
  P.~Roushan, N.~C. Rubin, D.~Sank, K.~J. Satzinger, V.~Smelyanskiy, K.~J.
  Sung, M.~D. Trevithick, A.~Vainsencher, B.~Villalonga, T.~White, Z.~J. Yao,
  P.~Yeh, A.~Zalcman, H.~Neven, and J.~M. Martinis, ``Quantum supremacy using a
  programmable superconducting processor,'' \emph{Nature}, vol. 574, no. 7779,
  pp. 505--510, Oct. 2019. [Online]. Available:
  \url{https://doi.org/10.1038/s41586-019-1666-5}
\BIBentrySTDinterwordspacing

\bibitem{Bardin2019}
\BIBentryALTinterwordspacing
J.~C. Bardin, T.~White, M.~Giustina, K.~J. Satzinger, K.~Arya, P.~Roushan,
  B.~Chiaro, J.~Kelly, Z.~Chen, B.~Burkett, Y.~Chen, E.~Jeffrey, A.~Dunsworth,
  A.~Fowler, B.~Foxen, C.~Gidney, R.~Graff, P.~Klimov, J.~Mutus, M.~J. McEwen,
  M.~Neeley, C.~J. Neill, E.~Lucero, C.~Quintana, A.~Vainsencher, H.~Neven,
  J.~Martinis, T.~Huang, S.~Das, D.~T. Sank, O.~Naaman, A.~E. Megrant, and
  R.~Barends, ``Design and {Characterization} of a 28-nm {Bulk}-{CMOS}
  {Cryogenic} {Quantum} {Controller} {Dissipating} {Less} {Than} 2 {mW} at 3
  {K},'' \emph{IEEE Journal of Solid-State Circuits}, vol.~54, no.~11, pp.
  3043--3060, Nov. 2019. [Online]. Available:
  \url{https://ieeexplore.ieee.org/document/8865458/}
\BIBentrySTDinterwordspacing

\bibitem{boissonneault2009dispersive}
M.~Boissonneault, J.~M. Gambetta, and A.~Blais, ``Dispersive regime of circuit
  qed: Photon-dependent qubit dephasing and relaxation rates,'' \emph{Physical
  Review A}, vol.~79, no.~1, p. 013819, 2009.

\bibitem{bradley1969optimizing}
S.~D. Bradley, ``Optimizing a scheme for run length encoding,''
  \emph{Proceedings of the IEEE}, vol.~57, no.~1, pp. 108--109, 1969.

\bibitem{bukhari2002dct}
K.~Bukhari, G.~Kuzmanov, and S.~Vassiliadis, ``Dct and idct implementations on
  different fpga technologies,'' in \emph{Proceedings of ProRISC}.\hskip 1em
  plus 0.5em minus 0.4em\relax Citeseer, 2002, pp. 232--235.

\bibitem{Butko2020}
\BIBentryALTinterwordspacing
A.~Butko, G.~Michelogiannakis, S.~Williams, C.~Iancu, D.~Donofrio, J.~Shalf,
  J.~Carter, and I.~Siddiqi, ``Understanding quantum control processor
  capabilities and limitations through circuit characterization,'' in
  \emph{2020 International Conference on Rebooting Computing ({ICRC})}.\hskip
  1em plus 0.5em minus 0.4em\relax {IEEE}, Dec. 2020. [Online]. Available:
  \url{https://doi.org/10.1109/icrc2020.2020.00011}
\BIBentrySTDinterwordspacing

\bibitem{chatterjee2018optimized}
S.~Chatterjee and K.~Sarawadekar, ``An optimized architecture of hevc core
  transform using real-valued dct coefficients,'' \emph{IEEE Transactions on
  Circuits and Systems II: Express Briefs}, vol.~65, no.~12, pp. 2052--2056,
  2018.

\bibitem{chen2012cacti}
K.~Chen, S.~Li, N.~Muralimanohar, J.~H. Ahn, J.~B. Brockman, and N.~P. Jouppi,
  ``Cacti-3dd: Architecture-level modeling for 3d die-stacked dram main
  memory,'' in \emph{2012 Design, Automation \& Test in Europe Conference \&
  Exhibition (DATE)}.\hskip 1em plus 0.5em minus 0.4em\relax IEEE, 2012, pp.
  33--38.

\bibitem{Chow2011}
\BIBentryALTinterwordspacing
J.~M. Chow, A.~D. C{\'{o}}rcoles, J.~M. Gambetta, C.~Rigetti, B.~R. Johnson,
  J.~A. Smolin, J.~R. Rozen, G.~A. Keefe, M.~B. Rothwell, M.~B. Ketchen, and
  M.~Steffen, ``Simple all-microwave entangling gate for fixed-frequency
  superconducting qubits,'' \emph{Physical Review Letters}, vol. 107, no.~8,
  Aug. 2011. [Online]. Available:
  \url{https://doi.org/10.1103/physrevlett.107.080502}
\BIBentrySTDinterwordspacing

\bibitem{clops}
``{Driving Quantum Performance},''
  \url{https://research.ibm.com/blog/circuit-layer-operations-per-second}.

\bibitem{das2020scalable}
P.~Das, C.~A. Pattison, S.~Manne, D.~Carmean, K.~Svore, M.~Qureshi, and
  N.~Delfosse, ``A scalable decoder micro-architecture for fault-tolerant
  quantum computing,'' 2020.

\bibitem{VanDijk2020_}
\BIBentryALTinterwordspacing
J.~P. G.~V. Dijk, B.~Patra, S.~Subramanian, X.~Xue, N.~Samkharadze, A.~Corna,
  C.~Jeon, F.~Sheikh, E.~Juarez-Hernandez, B.~P. Esparza, H.~Rampurawala, B.~R.
  Carlton, S.~Ravikumar, C.~Nieva, S.~Kim, H.-J. Lee, A.~Sammak, G.~Scappucci,
  M.~Veldhorst, L.~M.~K. Vandersypen, E.~Charbon, S.~Pellerano, M.~Babaie, and
  F.~Sebastiano, ``A scalable cryo-{CMOS} controller for the wideband
  frequency-multiplexed control of spin qubits and transmons,'' \emph{{IEEE}
  Journal of Solid-State Circuits}, vol.~55, no.~11, pp. 2930--2946, Nov. 2020.
  [Online]. Available: \url{https://doi.org/10.1109/jssc.2020.3024678}
\BIBentrySTDinterwordspacing

\bibitem{optimal4}
\BIBentryALTinterwordspacing
D.~J. Egger and F.~K. Wilhelm, ``Adaptive hybrid optimal quantum control for
  imprecisely characterized systems,'' \emph{Phys. Rev. Lett.}, vol. 112, p.
  240503, Jun 2014. [Online]. Available:
  \url{https://link.aps.org/doi/10.1103/PhysRevLett.112.240503}
\BIBentrySTDinterwordspacing

\bibitem{fang2002lightweight}
F.~Fang, T.~Chen, and R.~A. Rutenbar, ``Lightweight floating-point arithmetic:
  Case study of inverse discrete cosine transform,'' \emph{EURASIP Journal on
  Advances in Signal Processing}, vol. 2002, no.~9, pp. 1--14, 2002.

\bibitem{Fowler2012}
\BIBentryALTinterwordspacing
A.~G. Fowler, M.~Mariantoni, J.~M. Martinis, and A.~N. Cleland, ``Surface
  codes: Towards practical large-scale quantum computation,'' \emph{Phys. Rev.
  A}, vol.~86, p. 032324, Sep 2012. [Online]. Available:
  \url{https://link.aps.org/doi/10.1103/PhysRevA.86.032324}
\BIBentrySTDinterwordspacing

\bibitem{Frank2022}
\BIBentryALTinterwordspacing
D.~J. Frank, S.~Chakraborty, K.~Tien, P.~Rosno, T.~Fox, M.~Yeck, J.~A. Glick,
  R.~Robertazzi, R.~Richetta, J.~F. Bulzacchelli, D.~Ramirez, D.~Yilma,
  A.~Davies, R.~V. Joshi, S.~D. Chambers, S.~Lekuch, K.~Inoue, D.~Underwood,
  D.~Wisnieff, C.~Baks, D.~Bethune, J.~Timmerwilke, B.~R. Johnson, B.~P.
  Gaucher, and D.~J. Friedman, ``A cryo-{CMOS} low-power semi-autonomous qubit
  state controller in 14nm {FinFET} technology,'' in \emph{2022 {IEEE}
  International Solid- State Circuits Conference ({ISSCC})}.\hskip 1em plus
  0.5em minus 0.4em\relax {IEEE}, Feb. 2022. [Online]. Available:
  \url{https://doi.org/10.1109/isscc42614.2022.9731538}
\BIBentrySTDinterwordspacing

\bibitem{Fu2016}
\BIBentryALTinterwordspacing
X.~Fu, L.~Riesebos, L.~Lao, C.~G. Almudever, F.~Sebastiano, R.~Versluis,
  E.~Charbon, and K.~Bertels, ``\BIBforeignlanguage{en}{A heterogeneous quantum
  computer architecture},'' in \emph{\BIBforeignlanguage{en}{Proceedings of the
  {ACM} {International} {Conference} on {Computing} {Frontiers}}}.\hskip 1em
  plus 0.5em minus 0.4em\relax Como Italy: ACM, May 2016, pp. 323--330.
  [Online]. Available: \url{https://dl.acm.org/doi/10.1145/2903150.2906827}
\BIBentrySTDinterwordspacing

\bibitem{Fu2019}
X.~Fu, L.~Riesebos, M.~A. Rol, J.~van Straten, J.~van Someren, N.~Khammassi,
  I.~Ashraf, R.~F.~L. Vermeulen, V.~Newsum, K.~K.~L. Loh, J.~C. de~Sterke,
  W.~J. Vlothuizen, R.~N. Schouten, C.~G. Almudever, L.~DiCarlo, and
  K.~Bertels, ``eqasm: An executable quantum instruction set architecture,'' in
  \emph{2019 IEEE International Symposium on High Performance Computer
  Architecture (HPCA)}, 2019, pp. 224--237.

\bibitem{Fu2017}
\BIBentryALTinterwordspacing
X.~Fu, M.~A. Rol, C.~C. Bultink, J.~van Someren, N.~Khammassi, I.~Ashraf,
  R.~F.~L. Vermeulen, J.~C. de~Sterke, W.~J. Vlothuizen, R.~N. Schouten, C.~G.
  Almudever, L.~DiCarlo, and K.~Bertels, ``An experimental microarchitecture
  for a superconducting quantum processor,'' in \emph{Proceedings of the 50th
  {Annual} {IEEE}/{ACM} {International} {Symposium} on {Microarchitecture}},
  ser. {MICRO}-50 '17.\hskip 1em plus 0.5em minus 0.4em\relax Cambridge,
  Massachusetts: Association for Computing Machinery, Oct. 2017, pp. 813--825.
  [Online]. Available: \url{https://doi.org/10.1145/3123939.3123952}
\BIBentrySTDinterwordspacing

\bibitem{gebauerflexible}
R.~Gebauer, ``A flexible fpga-based control platform for superconducting
  multi-qubit experiments.''

\bibitem{weber}
``{Quantum Computer Datasheet},''
  \url{https://quantumai.google/hardware/datasheet/weber.pdf}, [Online;
  accessed 06-15-2021].

\bibitem{hashim2021randomized}
A.~Hashim, R.~K. Naik, A.~Morvan, J.-L. Ville, B.~Mitchell, J.~M. Kreikebaum,
  M.~Davis, E.~Smith, C.~Iancu, K.~P. O'Brien, I.~Hincks, J.~J. Wallman,
  J.~Emerson, and I.~Siddiqi, ``Randomized compiling for scalable quantum
  computing on a noisy superconducting quantum processor,'' 2021.

\bibitem{ibm_100q}
``{IBM Quantum breaks the 100‑qubit processor barrier},''
  \url{https://research.ibm.com/blog/127-qubit-quantum-processor-eagle}.

\bibitem{ibm_systems}
``{IBM Quantum Services},''
  \url{https://quantum-computing.ibm.com/services?services=systems}.

\bibitem{digiQ}
\BIBentryALTinterwordspacing
M.~R. Jokar, R.~Rines, G.~Pasandi, H.~Cong, A.~Holmes, Y.~Shi, M.~Pedram, and
  F.~T. Chong, ``Digiq: A scalable digital controller for quantum computers
  using sfq logic,'' 2022. [Online]. Available:
  \url{https://arxiv.org/abs/2202.01407}
\BIBentrySTDinterwordspacing

\bibitem{kang2021}
K.~Kang, B.~Kim, G.~Choi, S.-K. Lee, J.~Choi, J.~Lee, S.~Kang, M.~Lee, H.-J.
  Song, Y.~Chong, and J.-Y. Sim, ``A 5.5mw/channel 2-to-7 ghz frequency
  synthesizable qubit-controlling cryogenic pulse modulator for scalable
  quantum computers,'' in \emph{2021 Symposium on VLSI Circuits}, 2021, pp.
  1--2.

\bibitem{Kang2022}
\BIBentryALTinterwordspacing
K.~Kang, D.~Minn, S.~Bae, J.~Lee, S.~Bae, G.~Jung, S.~Kang, M.~Lee, H.-J. Song,
  and J.-Y. Sim, ``A cryo-{CMOS} controller {IC} with fully integrated
  frequency generators for superconducting qubits,'' in \emph{2022 {IEEE}
  International Solid- State Circuits Conference ({ISSCC})}.\hskip 1em plus
  0.5em minus 0.4em\relax {IEEE}, Feb. 2022. [Online]. Available:
  \url{https://doi.org/10.1109/isscc42614.2022.9731574}
\BIBentrySTDinterwordspacing

\bibitem{Kiene2021}
\BIBentryALTinterwordspacing
G.~Kiene, A.~Catania, R.~Overwater, P.~Bruschi, E.~Charbon, M.~Babaie, and
  F.~Sebastiano, ``13.4 a 1gs/s 6-to-8b 0.5mw/qubit cryo-{CMOS} {SAR} {ADC} for
  quantum computing in 40nm {CMOS},'' in \emph{2021 {IEEE} International Solid-
  State Circuits Conference ({ISSCC})}.\hskip 1em plus 0.5em minus 0.4em\relax
  {IEEE}, Feb. 2021. [Online]. Available:
  \url{https://doi.org/10.1109/isscc42613.2021.9365927}
\BIBentrySTDinterwordspacing

\bibitem{kim2022high}
Y.~Kim, A.~Morvan, L.~B. Nguyen, R.~K. Naik, C.~J{\"u}nger, L.~Chen, J.~M.
  Kreikebaum, D.~I. Santiago, and I.~Siddiqi, ``High-fidelity three-qubit
  itoffoli gate for fixed-frequency superconducting qubits,'' \emph{Nature
  Physics}, pp. 1--6, 2022.

\bibitem{Krantz2019}
\BIBentryALTinterwordspacing
P.~Krantz, M.~Kjaergaard, F.~Yan, T.~P. Orlando, S.~Gustavsson, and W.~D.
  Oliver, ``A quantum engineer{\textquotesingle}s guide to superconducting
  qubits,'' \emph{Applied Physics Reviews}, vol.~6, no.~2, p. 021318, Jun.
  2019. [Online]. Available: \url{https://doi.org/10.1063/1.5089550}
\BIBentrySTDinterwordspacing

\bibitem{krinner2021realizing}
S.~Krinner, N.~Lacroix, A.~Remm, A.~Di~Paolo, E.~Genois, C.~Leroux,
  C.~Hellings, S.~Lazar, F.~Swiadek, J.~Herrmann \emph{et~al.}, ``Realizing
  repeated quantum error correction in a distance-three surface code,''
  \emph{arXiv preprint arXiv:2112.03708}, 2021.

\bibitem{Lao2021}
L.~Lao, P.~Murali, M.~Martonosi, and D.~Browne, ``Designing calibration and
  expressivity-efficient instruction sets for quantum computing,'' in
  \emph{2021 ACM/IEEE 48th Annual International Symposium on Computer
  Architecture (ISCA)}, 2021, pp. 846--859.

\bibitem{leonard2019}
\BIBentryALTinterwordspacing
E.~Leonard, M.~A. Beck, J.~Nelson, B.~Christensen, T.~Thorbeck, C.~Howington,
  A.~Opremcak, I.~Pechenezhskiy, K.~Dodge, N.~Dupuis, M.~Hutchings, J.~Ku,
  F.~Schlenker, J.~Suttle, C.~Wilen, S.~Zhu, M.~Vavilov, B.~Plourde, and
  R.~McDermott, ``Digital coherent control of a superconducting qubit,''
  \emph{Phys. Rev. Applied}, vol.~11, p. 014009, Jan 2019. [Online]. Available:
  \url{https://link.aps.org/doi/10.1103/PhysRevApplied.11.014009}
\BIBentrySTDinterwordspacing

\bibitem{li2021qasmbench}
A.~Li, S.~Stein, S.~Krishnamoorthy, and J.~Ang, ``Qasmbench: A low-level qasm
  benchmark suite for nisq evaluation and simulation,'' 2021.

\bibitem{Li2019}
\BIBentryALTinterwordspacing
K.~Li, R.~McDermott, and M.~G. Vavilov, ``Hardware-efficient qubit control with
  single-flux-quantum pulse sequences,'' \emph{Physical Review Applied},
  vol.~12, no.~1, Jul. 2019. [Online]. Available:
  \url{https://doi.org/10.1103/physrevapplied.12.014044}
\BIBentrySTDinterwordspacing

\bibitem{liebermann2016}
\BIBentryALTinterwordspacing
P.~J. Liebermann and F.~K. Wilhelm, ``Optimal qubit control using single-flux
  quantum pulses,'' \emph{Phys. Rev. Applied}, vol.~6, p. 024022, Aug 2016.
  [Online]. Available:
  \url{https://link.aps.org/doi/10.1103/PhysRevApplied.6.024022}
\BIBentrySTDinterwordspacing

\bibitem{loeffler1989practical}
C.~Loeffler, A.~Ligtenberg, and G.~S. Moschytz, ``Practical fast 1-d dct
  algorithms with 11 multiplications,'' in \emph{International Conference on
  Acoustics, Speech, and Signal Processing,}.\hskip 1em plus 0.5em minus
  0.4em\relax IEEE, 1989, pp. 988--991.

\bibitem{lubinski2021application}
T.~Lubinski, S.~Johri, P.~Varosy, J.~Coleman, L.~Zhao, J.~Necaise, C.~H.
  Baldwin, K.~Mayer, and T.~Proctor, ``Application-oriented performance
  benchmarks for quantum computing,'' \emph{arXiv preprint arXiv:2110.03137},
  2021.

\bibitem{Magesan_2011}
\BIBentryALTinterwordspacing
E.~Magesan, J.~M. Gambetta, and J.~Emerson, ``Scalable and robust randomized
  benchmarking of quantum processes,'' \emph{Physical Review Letters}, vol.
  106, no.~18, May 2011. [Online]. Available:
  \url{http://dx.doi.org/10.1103/PhysRevLett.106.180504}
\BIBentrySTDinterwordspacing

\bibitem{marques2022logical}
J.~Marques, B.~Varbanov, M.~Moreira, H.~Ali, N.~Muthusubramanian,
  C.~Zachariadis, F.~Battistel, M.~Beekman, N.~Haider, W.~Vlothuizen
  \emph{et~al.}, ``Logical-qubit operations in an error-detecting surface
  code,'' \emph{Nature Physics}, vol.~18, no.~1, pp. 80--86, 2022.

\bibitem{mcdermott2014}
\BIBentryALTinterwordspacing
R.~McDermott and M.~G. Vavilov, ``Accurate qubit control with single flux
  quantum pulses,'' \emph{Phys. Rev. Applied}, vol.~2, p. 014007, Jul 2014.
  [Online]. Available:
  \url{https://link.aps.org/doi/10.1103/PhysRevApplied.2.014007}
\BIBentrySTDinterwordspacing

\bibitem{McDermott2018}
\BIBentryALTinterwordspacing
R.~McDermott, M.~G. Vavilov, B.~L.~T. Plourde, F.~K. Wilhelm, P.~J. Liebermann,
  O.~A. Mukhanov, and T.~A. Ohki, ``Quantum{\textendash}classical interface
  based on single flux quantum digital logic,'' \emph{Quantum Science and
  Technology}, vol.~3, no.~2, p. 024004, Jan. 2018. [Online]. Available:
  \url{https://doi.org/10.1088/2058-9565/aaa3a0}
\BIBentrySTDinterwordspacing

\bibitem{McEwen2021}
\BIBentryALTinterwordspacing
M.~McEwen, D.~Kafri, Z.~Chen, J.~Atalaya, K.~J. Satzinger, C.~Quintana, P.~V.
  Klimov, D.~Sank, C.~Gidney, A.~G. Fowler, F.~Arute, K.~Arya, B.~Buckley,
  B.~Burkett, N.~Bushnell, B.~Chiaro, R.~Collins, S.~Demura, A.~Dunsworth,
  C.~Erickson, B.~Foxen, M.~Giustina, T.~Huang, S.~Hong, E.~Jeffrey, S.~Kim,
  K.~Kechedzhi, F.~Kostritsa, P.~Laptev, A.~Megrant, X.~Mi, J.~Mutus,
  O.~Naaman, M.~Neeley, C.~Neill, M.~Niu, A.~Paler, N.~Redd, P.~Roushan, T.~C.
  White, J.~Yao, P.~Yeh, A.~Zalcman, Y.~Chen, V.~N. Smelyanskiy, J.~M.
  Martinis, H.~Neven, J.~Kelly, A.~N. Korotkov, A.~G. Petukhov, and R.~Barends,
  ``Removing leakage-induced correlated errors in superconducting quantum error
  correction,'' \emph{Nature Communications}, vol.~12, no.~1, Mar. 2021.
  [Online]. Available: \url{https://doi.org/10.1038/s41467-021-21982-y}
\BIBentrySTDinterwordspacing

\bibitem{McKay2017}
\BIBentryALTinterwordspacing
D.~C. McKay, C.~J. Wood, S.~Sheldon, J.~M. Chow, and J.~M. Gambetta,
  ``Efficient {Z}-{Gates} for {Quantum} {Computing},'' \emph{Physical Review
  A}, vol.~96, no.~2, p. 022330, Aug. 2017, arXiv: 1612.00858 version: 2.
  [Online]. Available: \url{http://arxiv.org/abs/1612.00858}
\BIBentrySTDinterwordspacing

\bibitem{VanMeter2013}
\BIBentryALTinterwordspacing
R.~V. Meter and C.~Horsman, ``A blueprint for building a quantum computer,''
  \emph{Communications of the {ACM}}, vol.~56, no.~10, pp. 84--93, Oct. 2013.
  [Online]. Available: \url{https://doi.org/10.1145/2494568}
\BIBentrySTDinterwordspacing

\bibitem{Meter2006}
\BIBentryALTinterwordspacing
R.~V. Meter and M.~Oskin, ``Architectural implications of quantum computing
  technologies,'' \emph{{ACM} Journal on Emerging Technologies in Computing
  Systems}, vol.~2, no.~1, pp. 31--63, Jan. 2006. [Online]. Available:
  \url{https://doi.org/10.1145/1126257.1126259}
\BIBentrySTDinterwordspacing

\bibitem{Metodi2011}
\BIBentryALTinterwordspacing
T.~S. Metodi, A.~I. Faruque, and F.~T. Chong, ``Quantum computing for computer
  architects, second edition,'' \emph{Synthesis Lectures on Computer
  Architecture}, vol.~6, no.~1, pp. 1--203, Mar. 2011. [Online]. Available:
  \url{https://doi.org/10.2200/s00331ed1v01y201101cac013}
\BIBentrySTDinterwordspacing

\bibitem{Nielsen2000-nt}
M.~A. Nielsen and I.~L. Chuang, \emph{\BIBforeignlanguage{en}{Cambridge series
  on information and the natural sciences: Quantum computation and quantum
  information}}.\hskip 1em plus 0.5em minus 0.4em\relax Cambridge, England:
  Cambridge University Press, Oct. 2000.

\bibitem{9365762}
J.-S. Park, S.~Subramanian, L.~Lampert, T.~Mladenov, I.~Klotchkov, D.~J.
  Kurian, E.~Juarez-Hernandez, B.~Perez-Esparza, S.~R. Kale, K.~T. Asma~Beevi,
  S.~Premaratne, T.~Watson, S.~Suzuki, M.~Rahman, J.~B. Timbadiya, S.~Soni, and
  S.~Pellerano, ``13.1 a fully integrated cryo-cmos soc for qubit control in
  quantum computers capable of state manipulation, readout and high-speed gate
  pulsing of spin qubits in intel 22nm ffl finfet technology,'' in \emph{2021
  IEEE International Solid- State Circuits Conference (ISSCC)}, vol.~64, 2021,
  pp. 208--210.

\bibitem{icarus}
\BIBentryALTinterwordspacing
K.~H. Park, Y.~S. Yap, Y.~P. Tan, C.~Hufnagel, L.~H. Nguyen, K.~H. Lau,
  S.~Efthymiou, S.~Carrazza, R.~P. Budoyo, and R.~Dumke, ``Icarus-q: A scalable
  rfsoc-based control system for superconducting quantum computers,'' 2021.
  [Online]. Available: \url{https://arxiv.org/abs/2112.02933}
\BIBentrySTDinterwordspacing

\bibitem{Pauka2021}
\BIBentryALTinterwordspacing
S.~J. Pauka, K.~Das, R.~Kalra, A.~Moini, Y.~Yang, M.~Trainer, A.~Bousquet,
  C.~Cantaloube, N.~Dick, G.~C. Gardner, M.~J. Manfra, and D.~J. Reilly,
  ``\BIBforeignlanguage{en}{A cryogenic {CMOS} chip for generating control
  signals for multiple qubits},'' \emph{\BIBforeignlanguage{en}{Nature
  Electronics}}, vol.~4, no.~1, pp. 64--70, Jan. 2021. [Online]. Available:
  \url{http://www.nature.com/articles/s41928-020-00528-y}
\BIBentrySTDinterwordspacing

\bibitem{pekhimenko2012base}
G.~Pekhimenko, V.~Seshadri, O.~Mutlu, M.~A. Kozuch, P.~B. Gibbons, and T.~C.
  Mowry, ``Base-delta-immediate compression: Practical data compression for
  on-chip caches,'' in \emph{2012 21st international conference on parallel
  architectures and compilation techniques (PACT)}.\hskip 1em plus 0.5em minus
  0.4em\relax IEEE, 2012, pp. 377--388.

\bibitem{poremba2015destiny}
M.~Poremba, S.~Mittal, D.~Li, J.~S. Vetter, and Y.~Xie, ``Destiny: A tool for
  modeling emerging 3d nvm and edram caches,'' in \emph{2015 Design, Automation
  \& Test in Europe Conference \& Exhibition (DATE)}.\hskip 1em plus 0.5em
  minus 0.4em\relax IEEE, 2015, pp. 1543--1546.

\bibitem{propson2022robust}
T.~Propson, B.~E. Jackson, J.~Koch, Z.~Manchester, and D.~I. Schuster, ``Robust
  quantum optimal control with trajectory optimization,'' \emph{Physical Review
  Applied}, vol.~17, no.~1, p. 014036, 2022.

\bibitem{recommendation199281}
T.~Recommendation, ``81: Digital compression and coding of continuous-tone
  still images—requirements and guidelines,'' Technical report, CCITT, Tech.
  Rep., 1992.

\bibitem{Rigetti2010}
\BIBentryALTinterwordspacing
C.~Rigetti and M.~Devoret, ``Fully microwave-tunable universal gates in
  superconducting qubits with linear couplings and fixed transition
  frequencies,'' \emph{Physical Review B}, vol.~81, no.~13, Apr. 2010.
  [Online]. Available: \url{https://doi.org/10.1103/physrevb.81.134507}
\BIBentrySTDinterwordspacing

\bibitem{ibm1000}
``{IBM Quantum Roadmap},''
  \url{https://www.ibm.com/blogs/research/2020/09/ibm-quantum-roadmap/}.

\bibitem{optimal2}
\BIBentryALTinterwordspacing
K.~Rojan, D.~M. Reich, I.~Dotsenko, J.-M. Raimond, C.~P. Koch, and G.~Morigi,
  ``Arbitrary-quantum-state preparation of a harmonic oscillator via optimal
  control,'' \emph{Phys. Rev. A}, vol.~90, p. 023824, Aug 2014. [Online].
  Available: \url{https://link.aps.org/doi/10.1103/PhysRevA.90.023824}
\BIBentrySTDinterwordspacing

\bibitem{rol2019fast}
M.~Rol, F.~Battistel, F.~Malinowski, C.~Bultink, B.~Tarasinski, R.~Vollmer,
  N.~Haider, N.~Muthusubramanian, A.~Bruno, B.~Terhal \emph{et~al.}, ``Fast,
  high-fidelity conditional-phase gate exploiting leakage interference in
  weakly anharmonic superconducting qubits,'' \emph{Physical review letters},
  vol. 123, no.~12, p. 120502, 2019.

\bibitem{scipy}
``Scipy fftpack,''
  \url{https://docs.scipy.org/doc/scipy/reference/generated/scipy.fftpack.dct.html}.

\bibitem{shen2012unified}
S.~Shen, W.~Shen, Y.~Fan, and X.~Zeng, ``A unified 4/8/16/32-point integer idct
  architecture for multiple video coding standards,'' in \emph{2012 IEEE
  International Conference on Multimedia and Expo}.\hskip 1em plus 0.5em minus
  0.4em\relax IEEE, 2012, pp. 788--793.

\bibitem{Shi2019}
\BIBentryALTinterwordspacing
Y.~Shi, N.~Leung, P.~Gokhale, Z.~Rossi, D.~I. Schuster, H.~Hoffmann, and F.~T.
  Chong, ``Optimized compilation of aggregated instructions for realistic
  quantum computers,'' in \emph{Proceedings of the Twenty-Fourth International
  Conference on Architectural Support for Programming Languages and Operating
  Systems}.\hskip 1em plus 0.5em minus 0.4em\relax {ACM}, Apr. 2019. [Online].
  Available: \url{https://doi.org/10.1145/3297858.3304018}
\BIBentrySTDinterwordspacing

\bibitem{singhadia2020hardware}
A.~Singhadia, M.~Mamillapalli, and I.~Chakrabarti, ``Hardware-efficient
  2d-dct/idct architecture for portable hevc-compliant devices,'' \emph{IEEE
  Transactions on Consumer Electronics}, vol.~66, no.~3, pp. 203--212, 2020.

\bibitem{optimal3}
\BIBentryALTinterwordspacing
A.~Sp\"orl, T.~Schulte-Herbr\"uggen, S.~J. Glaser, V.~Bergholm, M.~J. Storcz,
  J.~Ferber, and F.~K. Wilhelm, ``Optimal control of coupled josephson
  qubits,'' \emph{Phys. Rev. A}, vol.~75, p. 012302, Jan 2007. [Online].
  Available: \url{https://link.aps.org/doi/10.1103/PhysRevA.75.012302}
\BIBentrySTDinterwordspacing

\bibitem{qick}
\BIBentryALTinterwordspacing
L.~Stefanazzi, K.~Treptow, N.~Wilcer, C.~Stoughton, S.~Montella, C.~Bradford,
  G.~Cancelo, S.~Saxena, H.~Arnaldi, S.~Sussman, A.~Houck, A.~Agrawal,
  H.~Zhang, C.~Ding, and D.~I. Schuster, ``The qick (quantum instrumentation
  control kit): Readout and control for qubits and detectors,'' 2021. [Online].
  Available: \url{https://arxiv.org/abs/2110.00557}
\BIBentrySTDinterwordspacing

\bibitem{Stojanovi2012}
\BIBentryALTinterwordspacing
V.~M. Stojanovi{\'{c}}, A.~Fedorov, A.~Wallraff, and C.~Bruder,
  ``Quantum-control approach to realizing a toffoli gate in circuit {QED},''
  \emph{Physical Review B}, vol.~85, no.~5, Feb. 2012. [Online]. Available:
  \url{https://doi.org/10.1103/physrevb.85.054504}
\BIBentrySTDinterwordspacing

\bibitem{sze2014high}
V.~Sze, M.~Budagavi, and G.~J. Sullivan, ``High efficiency video coding
  (hevc),'' in \emph{Integrated circuit and systems, algorithms and
  architectures}.\hskip 1em plus 0.5em minus 0.4em\relax Springer, 2014,
  vol.~39, p.~40.

\bibitem{tannu2017cryogenic}
S.~S. Tannu, D.~M. Carmean, and M.~K. Qureshi, ``Cryogenic-dram based memory
  system for scalable quantum computers: A feasibility study,'' in
  \emph{Proceedings of the International Symposium on Memory Systems}, 2017,
  pp. 189--195.

\bibitem{Tannu2017}
\BIBentryALTinterwordspacing
S.~S. Tannu, Z.~A. Myers, P.~J. Nair, D.~M. Carmean, and M.~K. Qureshi,
  ``Taming the instruction bandwidth of quantum computers via hardware-managed
  error correction,'' in \emph{Proceedings of the 50th Annual {IEEE}/{ACM}
  International Symposium on Microarchitecture}.\hskip 1em plus 0.5em minus
  0.4em\relax {ACM}, Oct. 2017. [Online]. Available:
  \url{https://doi.org/10.1145/3123939.3123940}
\BIBentrySTDinterwordspacing

\bibitem{Tomita2014}
\BIBentryALTinterwordspacing
Y.~Tomita and K.~M. Svore, ``Low-distance surface codes under realistic quantum
  noise,'' \emph{Phys. Rev. A}, vol.~90, p. 062320, Dec 2014. [Online].
  Available: \url{https://link.aps.org/doi/10.1103/PhysRevA.90.062320}
\BIBentrySTDinterwordspacing

\bibitem{tran1999fast}
T.~D. Tran, ``Fast multiplierless approximation of the dct,'' \emph{Synthesis},
  vol.~10, p.~1, 1999.

\bibitem{vanDijk2020}
\BIBentryALTinterwordspacing
J.~P.~G. van Dijk, B.~Patra, S.~Pellerano, E.~Charbon, F.~Sebastiano, and
  M.~Babaie, ``Designing a {DDS}-based {SoC} for high-fidelity multi-qubit
  control,'' \emph{{IEEE} Transactions on Circuits and Systems I: Regular
  Papers}, vol.~67, no.~12, pp. 5380--5393, Dec. 2020. [Online]. Available:
  \url{https://doi.org/10.1109/tcsi.2020.3019413}
\BIBentrySTDinterwordspacing

\bibitem{Xie2022}
\BIBentryALTinterwordspacing
L.~Xie, J.~Zhai, Z.~Zhang, J.~Allcock, S.~Zhang, and Y.-C. Zheng, ``Suppressing
  {ZZ} crosstalk of quantum computers through pulse and scheduling
  co-optimization,'' in \emph{Proceedings of the 27th {ACM} International
  Conference on Architectural Support for Programming Languages and Operating
  Systems}.\hskip 1em plus 0.5em minus 0.4em\relax {ACM}, Feb. 2022. [Online].
  Available: \url{https://doi.org/10.1145/3503222.3507761}
\BIBentrySTDinterwordspacing

\bibitem{xilinx}
``{Xilinx Ultrascale},''
  \url{https://docs.xilinx.com/v/u/en-US/ds890-ultrascale-overview}.

\bibitem{Xue2021}
\BIBentryALTinterwordspacing
X.~Xue, B.~Patra, J.~P.~G. van Dijk, N.~Samkharadze, S.~Subramanian, A.~Corna,
  B.~P. Wuetz, C.~Jeon, F.~Sheikh, E.~Juarez-Hernandez, B.~P. Esparza,
  H.~Rampurawala, B.~Carlton, S.~Ravikumar, C.~Nieva, S.~Kim, H.-J. Lee,
  A.~Sammak, G.~Scappucci, M.~Veldhorst, F.~Sebastiano, M.~Babaie,
  S.~Pellerano, E.~Charbon, and L.~M.~K. Vandersypen, ``{CMOS}-based cryogenic
  control of silicon quantum circuits,'' \emph{Nature}, vol. 593, no. 7858, pp.
  205--210, May 2021. [Online]. Available:
  \url{https://doi.org/10.1038/s41586-021-03469-4}
\BIBentrySTDinterwordspacing

\bibitem{zahedinejad2016designing}
E.~Zahedinejad, J.~Ghosh, and B.~C. Sanders, ``Designing high-fidelity
  single-shot three-qubit gates: a machine-learning approach,'' \emph{Physical
  Review Applied}, vol.~6, no.~5, p. 054005, 2016.

\bibitem{zhao2021realizing}
Y.~Zhao, Y.~Ye, H.-L. Huang, Y.~Zhang, D.~Wu, H.~Guan, Q.~Zhu, Z.~Wei, T.~He,
  S.~Cao \emph{et~al.}, ``Realizing an error-correcting surface code with
  superconducting qubits,'' \emph{arXiv preprint arXiv:2112.13505}, 2021.

\bibitem{zhou2022effective}
Z.~Zhou and Z.~Pan, ``Effective hardware accelerator for 2d dct/idct using
  improved loeffler architecture,'' \emph{IEEE Access}, vol.~10, pp.
  11\,011--11\,020, 2022.

\end{thebibliography}

\end{document}